\documentclass[twocolumn,pra,superscriptaddress,showpacs,eqsecnum,twoside]{revtex4}

\usepackage[dvips]{graphicx,color}%
\usepackage{dcolumn}
\usepackage{amsmath,amssymb} 

\makeatletter
\def\btt#1{\texttt{\@backslashchar#1}}%
\DeclareRobustCommand\bblash{\btt{\@backslashchar}}%
\makeatother

\begin{document}


\title[Dicke-Type Energy Level 
Crossings in 
Cavity-Induced Atom Cooling]{Dicke-Type Energy Level 
Crossings in Cavity-Induced Atom Cooling: \\  
Another Superradiant Cooling 
}
\author{Masao Hirokawa}
\email{hirokawa@math.okayama-u.ac.jp}
\homepage{http://www.math.okayama-u.ac.jp/~hirokawa}
\affiliation{%
Department of Mathematics, Okayama University, 
Okayama 700-8530, Japan  
}%

\date{\today}

\begin{abstract}
This paper is devoted to energy-spectral analysis for the 
system of a two-level atom coupled with photons in a cavity. 
It is shown that the Dicke-type energy level crossings 
take place when the atom-cavity interaction of 
the system undergoes changes 
between the weak coupling regime 
and the strong one. 
Using the phenomenon of the crossings 
we develop the idea of cavity-induced 
atom cooling proposed by the group of Ritsch, 
and we lay mathematical foundations 
of a possible mechanism for 
another superradiant cooling 
in addition to that proposed by Domokos and Ritsch. 
The process of our superradiant 
cooling can function well by cavity decay and 
by control of the position of the atom, 
at least in (mathematical) theory, 
even if there is neither atomic absorption 
nor atomic emission of photons.  
\end{abstract}

\pacs{37.10.De,42.50.Pq,02.30.Sa,02.30.Tb}
\maketitle
\pagestyle{myheadings}
 \markboth{M. Hirokawa}{Dicke-Type Energy Level 
Crossings in Cavity-Induced Atom Cooling}


\section{Introduction}
\label{sec:intro}

Laser cooling is one of attractive subjects of 
modern physics. 
It has been demonstrated with several experimental 
techniques such as the ion cooling \cite{NHTD}, 
the Doppler cooling \cite{CHBCA}, 
the Sisyphus cooling \cite{C-TP}, 
etc. 
Also it has enabled us to observe 
many fundamental phenomena in 
theoretical physics. 
One of typical instances of applications 
of the laser cooling is 
the observation of Bose-Einstein 
condensation \cite{CW,ketterle}.
It has been about ten years 
since another type of laser cooling 
was proposed using the strong atom-photon 
interaction. 
Such a strong interaction between 
atom and photons is realized in the so-called 
cavity quantum electrodynamics (QED) 
\cite{HCLK,RBH,MD,MBBBK,dutra,HR06}. 
Thus, the group of Ritsch has 
investigated the system of 
a two-level atom coupled with a laser, 
and then, 
they have found a mechanism 
for cooling the atom  \cite{H-R}, 
which is similar to that of 
the Sisyphus cooling. 
The cooling mechanism is called 
\textit{cavity-induced atom cooling}. 
In the process of their cavity-induced 
atom cooling, 
the method to carry away the energy from 
the atom coupled with photons is given 
not only by atomic decay 
(i.e., atomic spontaneous emission of photons) 
but also by cavity decay. 
It has experimentally been confirmed 
that the cavity decay works in the cooling system 
\cite{H-K,F-R}. 
Concerning the cooling methods using 
cavity QED, 
Domokos and Ritsch proposed a concept 
of superradiant cooling \cite{DR} 
based on the atomic self-organization 
and cooperation among many atoms in a cavity \cite{DR02}. 
The atom-photon interaction in the strong coupling regime 
brings the situation amazingly different from ordinary 
atomic decay \cite{dutra}. 
We will adopt this difference 
into our arguments on the cavity-induced 
atom cooling. 

As well as the ensemble of 
many two-level atoms coupled with 
a laser has the possibility of making 
superradiance as a cooperative 
effect in optics \cite{GK,AEI}, 
another superradiance may also appear 
in energy spectrum 
even for the system of a two-level atom 
coupled with a laser, provided that it is in 
the strong coupling regime 
\cite{preparata,enz,PL,hir01,hir02}. 
As far as atom-laser interaction 
in the cooling process for the 
Bose-Einstein condensation goes, 
a superradiance has been experimentally observed 
under a certain physical condition 
\cite{S-K99,I-K,S-K,F-B}, and this phenomenon has 
been theoretically shown \cite{PVZ,PVZ04,PVZ05}. 
It was pointed out that 
there is a possibility that superradiance causes 
the energy level crossing between the initial 
ground state energy and an initial excited state 
energy \cite{PL,hir01,hir02}, namely, a kind of 
phase transition occurs. 
This is an optical phenomenon of light-induced 
phase transitions though it is not a cooperative 
effect in optics. 
The details of such an energy level crossing 
has precisely been studied, 
and then, 
this type of crossing is called 
the \textit{Dicke-type} (\textit{energy 
level}) \textit{crossing} \cite{hir-iumj}. 
We will strictly define its meaning 
in Sec.\ref{sec:Hamiltonian}. 
The reason why we call the energy level 
crossing so is that it is basically caused by 
the mathematical mechanism \cite{PL,hir02} of 
Dicke's superradiance \cite{dicke}.

This paper is devoted to developing the 
cavity-induced atom cooling. 
Namely, we will show that the Dicke-type energy level 
crossings take place when the system undergoes 
changes between the weak coupling regime 
and the strong one. 
Using the crossings, 
we will propose a possibility of another superradiant 
cooling in terms of the energy spectrum 
from our point of view. 
In our proposal we will consider whether the followings 
are possible in theory for cooling the atom 
in a cavity: 
1) can we use a laser only for controlling the strength 
of the atom-cavity interaction without 
throwing another laser to the atom for 
driving it to an excited state? 
2) can we expect that the energy loss caused 
by cavity decay 
becomes much larger? 
To perform our research into the problems, 
we consider an ideal situation 
only to see the energy-spectral property 
for our system 
without considering, for example, 
the laser heating processes caused by 
diffusion of the atomic momentum. 
Some mathematical technique to 
make the spectral analysis for such systems 
have been developed 
lately \cite{hir-iumj,PW,NNW,parmeggiani,IW}. 
Thus, another purpose of this paper is to show 
the mechanism of the Dicke-type energy level crossing 
in the cavity-induced atom cooling 
as rigorously as in the works in Ref.\onlinecite{HL} 
so that the process of our superradiant 
cooling can function well 
in (mathematical) theory. 

Our paper is constructed in the following. 
In Sec.\ref{sec:Hamiltonian} 
we will generalize the Hamiltonian 
$H(\Omega,\alpha;d)$ 
which Ritsch's group handled, 
where $\Omega$ is a function of space-time point and 
governs the atom-photon interaction, 
$\alpha$ also a function of space-time 
and a generalization of the strength of 
the pump field, and $d$ a parameter for 
non-linear coupling of the atom and photons. 
Moreover, we will define some notion to explain 
what the Dicke-type energy level crossing is.
We will make energy-spectral analysis 
for the generalized Hamiltonian 
$H(\Omega,\alpha;d)$ 
in and after Sec.\ref{sec:dicke-type-crossing}. 
In Sec.\ref{sec:dicke-type-crossing} 
we will show that the Dicke-type 
energy level crossings take place 
for $H(\Omega,\alpha;d)$ 
with $\alpha\equiv 0$. 
In Sec.\ref{sec:sgse} 
we will show the existence of the 
superradiant ground state energy 
for $H(\Omega,\alpha;d)$ 
with $\alpha\equiv 0$ and $d=1$ in the strong 
coupling regime. 
In Sec.\ref{sec:stability} we will argue 
the stability of the Dicke-type energy level 
crossing under the condition 
$\alpha\equiv\!\!\!\!\!\!{/}\,\, 0$. 

\section{Hamiltonian and Some Notion}
\label{sec:Hamiltonian}

In Ref.\onlinecite{H-R} the group of Ritsch 
studied a Hamiltonian adopting dipole and rotating 
wave approximation. 
To write down their Hamiltonian, 
we define some operators: 
the atomic position 
(resp. momentum) operator is denoted by 
$x$ (resp. $p$), 
the photon annihilation (resp. creation) 
operator by $a$ (resp. $a^{\dagger}$), 
and the atomic operator is given by 
$\sigma_{ij} = |i\rangle\langle j|$, $i,j = 0, 1$. 
Then, the Hamiltonian is 
\begin{align*}
H =& 
\frac{1}{2m}p^{2} 
- \Delta\sigma_{11} 
- \Delta_{\mathrm{c}}a^{\dagger}a \\ 
&+ i\Omega(x)
\left(\sigma_{01}a^{\dagger} - \sigma_{10}a\right) 
+ i\alpha(a-a^{\dagger}), 
\end{align*}
where two real numbers $\Delta$ and $\Delta_{\mathrm{c}}$ 
with $-\infty < \Delta < +\infty$ and $\Delta_{\mathrm{c}}<0$ 
are respectively the atom-pump detuning and 
the detuning of the empty cavity relative to 
the pump frequency, 
and $\Omega(x)$ stands for the atom-cavity 
coupling constant, i.e., 
$\Omega(x) = \Omega_{0}\cos kx$ 
with the position $x$ of the atom and 
the wave number $k$ 
of photons of the laser. 
We note that in the case $\alpha=0$ 
the Hamiltonian $H$ is used to argue 
the resonant interaction of an atom 
with a microwave field 
\cite{SZ}. 
In Hamiltonian $H$, 
the part consisting of the first, the second, 
and the third terms (i.e., 
$(2m)^{-1}p^{2}- \Delta\sigma_{11} 
- \Delta_{\mathrm{c}}a^{\dagger}a$) 
is the free Hamiltonian of our system. 
Each of the fourth and fifth terms 
represents the Hamiltonian of interaction and 
the energy operator of 
the pump field respectively. 
In this section we generalize the above Hamiltonian. 
Our generalization is the following: 
(1) we consider not only the linear coupling 
but also non-linear coupling; 
(2) we introduce the time-dependence into 
the coupling constant $\Omega(x)$; 
(3) we consider the general operator 
which represents not only the energy operator 
of the pump field but also, for instance, 
the energy operator of the pump field plus 
some error potential coming from 
the environment of the experiment for testing 
the system. 
Thus our Hamiltonian reads
\begin{align*}
H(\Omega,\alpha;d) =& 
\frac{1}{2m}p^{2} 
- \Delta\sigma_{11} 
- \Delta_{\mathrm{c}}a^{\dagger}a \notag \\ 
&+ i\Omega(x,t)
\left(\sigma_{01}a^{\dagger d} - \sigma_{10}a^{d}\right) 
+ \alpha(x,t)W(x,t)
\end{align*}
for $d = 1, 2, \cdots$, 
where $\Omega(x,t)$ and $\alpha(x,t)$ are continuous, 
real-valued functions of $(x,t)$ with 
$\Omega(x,0)=0=\alpha(x,0)$ for every 
position $x$ of the atom, 
and $\alpha(x,t)W(x,t)$ 
is the generalization of the energy operator 
of the pump field. 
As an example of $\Omega(x,t)$, 
we often adopt $\Omega(x,t)=\Omega_{0}(t)\gamma(x)$ 
in this paper. 
Here $\Omega_{0}(t)$ is a continuous, real-valued 
function of time $t \ge 0$ with $\Omega_{0}(0)=0$, 
and $\gamma(x)$ a bounded, continuous, real-valued 
function of the position $x$ of the atom. 
For instance, $\gamma(x) = \cos kx$.

In the case where $\Omega(x,t)\equiv 0$ 
and $\alpha(z,t)\equiv 0$, 
we denote eigenvalues of 
$H(0,0;d) := 
H(\Omega=0,\alpha=0;d)$ 
by $\mathcal{E}_{0} < \mathcal{E}_{1} 
< \cdots < \mathcal{E}_{n} < \cdots$. 
When either $\Omega(x,t)$ or $\alpha(t)$ is alive, 
we denote eigenvalues of 
$H(\Omega,\alpha;d)$ 
by $\mathcal{E}_{n}(\Omega,\alpha;d)$ 
for $n = 0, 1, \cdots$. 
If the interaction 
$H_{\mathrm{int}}:= i\Omega(t,x)
\left(\sigma_{01}a^{\dagger d} - \sigma_{10}a^{d}\right) 
+ \alpha(t)W(x,t)$ is a small perturbation 
for H(0,0;d), 
then each eigenvalue $\mathcal{E}_{n}(\Omega,\alpha;d)$ sits 
near its original 
position $\mathcal{E}_{n}$, 
so that the primary order among eigenvalues 
is kept: $\mathcal{E}_{0}(\Omega,\alpha;d) 
< \mathcal{E}_{1}(\Omega,\alpha;d) 
< \cdots < \mathcal{E}_{n}(\Omega,\alpha;d) < \cdots$.
On the other hand, 
the phase transition of the superradiance 
\cite{PL,hir01,hir02} tells us 
about a possibility that 
$\mathcal{E}_{1}(\Omega,\alpha;d)$ is less than 
$\mathcal{E}_{0}(\Omega,\alpha;d)$ and thus becomes 
a new ground state energy 
provided that the interaction 
$H_{\mathrm{int}}$ has some strong strength. 
For our Hamiltonian, we can classify crossings 
into two types. 
One type is the crossing between an ascending eigenvalue 
$\mathcal{E}_{n}^{+}(\Omega,\alpha;d)$ 
and a descending one 
$\mathcal{E}_{n}^{-}(\Omega,\alpha;d)$ 
as the strength of the interaction 
$H_\mathrm{int}$ grows enough. 
Another type is the crossing only among 
descending eigenvalues 
$\mathcal{E}_{n}^{-}(\Omega,\alpha;d)$ 
(or ascending eigenvalues 
$\mathcal{E}_{n}^{+}(\Omega,\alpha;d)$). 
We call the former type a \textit{trivial} crossing, 
and the latter type a \textit{non-trivial} crossing. 
For the non-trivial crossing, 
as the strength of the interaction becomes much stronger, 
even many $\mathcal{E}_{n}^{-}(\Omega,\alpha;d)$s may 
be less than $\mathcal{E}_{0}^{-}(\Omega,\alpha;d)$. 
We call such a non-trivial crossing 
the Dicke-type (energy level) crossing \cite{hir-iumj}. 
Moreover, $\mathcal{E}_{n}^{-}(\Omega,\alpha;d)$s 
are capable of 
usurping the position of the ground state energy 
in turn. 
We call such a new ground state energy 
the \textit{superradiant ground state energy}. 
The Dicke-type energy level crossings and 
the appearance of the superradiant ground state 
energy can be used, together with cavity decay, 
for carrying away the energy 
from the system. 
Based on this idea, 
we construct mathematical foundations 
of the concept of superradiant cooling 
different from that proposed in 
Ref.\onlinecite{DR} 
in and after the next section.

\section{The Dicke-Type Energy Level Crossings in The Case 
$\alpha\equiv 0$.} 
\label{sec:dicke-type-crossing}

In this section we show how the Dicke-type 
energy level crossing takes place for 
$H(\Omega,0;d)$, that is, for 
$H(\Omega,\alpha;d)$ in the case where 
$\alpha(t)\equiv 0$. 
As in Ref.\onlinecite{H-R} 
using the well-known identification 
so that the ground state $|0\rangle$ with 
the energy $\varepsilon_{0}$ 
and the $1$st excited state $|1\rangle$ 
with the energy $\varepsilon_{1}$ 
are unitarily equivalent to 
$\begin{pmatrix}
0 \\ 1 
\end{pmatrix}
$ 
and 
$\begin{pmatrix}
1 \\ 0 
\end{pmatrix}
$ 
respectively, 
$H(\Omega,0;d)$ approximately reads 
\begin{equation}
H_{0}(z,t;d)
:=\begin{pmatrix}
- \Delta_{\mathrm{c}}a^{\dagger}a + 
\varepsilon_{1} - \Delta 
& -i\Omega(z,t)a^{d} \\ 
i\Omega(z,t)a^{\dagger d} 
& - \Delta_{\mathrm{c}}a^{\dagger}a + 
\varepsilon_{0} 
\end{pmatrix}. 
\label{eq:JJ}  
\end{equation}
Here we note 
$\sigma_{00}
=
\begin{pmatrix}
0 & 0 \\ 
0 & 1 
\end{pmatrix}$, 
$\sigma_{01}
=
\begin{pmatrix}
0 & 0 \\ 
1 & 0 
\end{pmatrix}$, 
$\sigma_{10}
=
\begin{pmatrix}
0 & 1 \\ 
0 & 0 
\end{pmatrix}$, 
and 
$\sigma_{11}
=
\begin{pmatrix}
1 & 0 \\ 
0 & 0 
\end{pmatrix}$. 
In their paper we always assume that 
\begin{equation}
\varepsilon_{1}-\Delta > \varepsilon_{0}. 
\label{eq:assumption}
\end{equation}
Therefore, the ground state energy of 
$H_{0}(z,0;d)$ (i.e., $H_{0}(z,t;d)$ with 
$\Omega(z,t)\equiv 0$) is always $\varepsilon_{0}$.

As shown in Sec.\ref{sec:appendix} with the 
method which is a generalization 
of that in Refs.\onlinecite{hir01} 
and \onlinecite{BR}, 
since $H_{0}(z,t;d)$ is basically the 
Hamiltonian of the Jaynes-Cummings model 
\cite{JC},
all the energy levels of $H_{0}(z,t;d)$ 
are perfectly determined. 
They are given by 
$- \Delta_{\mathrm{c}}n+\varepsilon_{0}$ 
for non-negative integer 
$n$ with $n < d$, and $\Xi_{n}(d)\pm 
\Upsilon_{n}(z,t;d)$ 
for non-negative integer $n$ with $n \ge d$,
where 
$$
\Xi_{n}(d) 
= 
- \Delta_{\mathrm{c}}n 
+ \frac{1}{2}
\left(\varepsilon_{0}+\varepsilon_{1}
+d\Delta_{\mathrm{c}}-\Delta\right), 
$$
and $\Upsilon_{n}(z,t;d)$ is the generalized 
Rabi frequency \cite{ME}: 
\begin{align}
&\Upsilon_{n}(z,t;d) \notag \\ 
=& 
\frac{1}{2}\sqrt{
(\varepsilon_{1}-\varepsilon_{0}
+d\Delta_{\mathrm{c}}-\Delta)^{2}
+ 4|\Omega(z,t)|^{2}
\frac{n!}{(n-d)!}}\,\,\,.
\label{eq:GRF}
\end{align}
Here we note all the energy levels of 
$H_{0}(z,0;d)$ are 
$- \Delta_{\mathrm{c}}n + \varepsilon_{0}$ 
and $- \Delta_{\mathrm{c}}n + \varepsilon_{1}
- \Delta$, 
$n=0, 1, \cdots$. 
Therefore, we can conclude that 
the energy levels of $H_{0}(z,t;d)$ 
are completely given by energies 
$E_{n}^{0}(z,t;d)$ and energies 
$E_{n}^{\pm}(z,t;d)$ 
of the generalized 
Jaynes-Cummings doublet 
\cite{GK}, 
continuous functions of $(z,t)$, 
for each $n = 0, 1, \cdots$: 
For non-negative integers $n$ with $n < d$
$$
E_{n}^{0}(z,t;d) = -\Delta_{\mathrm{c}}n
+\varepsilon_{0}. 
$$
For non-negative integers $n$ with $n \ge d$, 
on the other hand, 
$$
E_{n}^{-}(z,t;d) 
= 
\begin{cases}
\Xi_{n}(d) - \Upsilon_{n}(z,t;d) 
& \quad \\ 
\qquad\qquad\qquad\qquad\text{if $
\varepsilon_{1}-\varepsilon_{0}\ge
\Delta-d\Delta_{\mathrm{c}}$}, 
&\quad \\ 
\Xi_{n+d}(d) - \Upsilon_{n+d}(z,t;d) 
& \quad \\ 
\qquad\qquad\qquad\qquad\text{if $
\varepsilon_{1}-\varepsilon_{0}<
\Delta-d\Delta_{\mathrm{c}}$}, 
\end{cases}
$$ 
and 
$$
E_{n}^{+}(z,t;d) 
= 
\begin{cases}
\Xi_{n+d}(d) + \Upsilon_{n+d}(z,t;d) 
& \quad \\ 
\qquad\qquad\qquad\qquad\text{if $
\varepsilon_{1}-\varepsilon_{0}\ge
\Delta-d\Delta_{\mathrm{c}}$}, 
& \quad \\ 
\Xi_{n}(d) + \Upsilon_{n}(z,t;d) 
& \quad \\ 
\qquad\qquad\qquad\qquad\text{if $
\varepsilon_{1}-\varepsilon_{0}<
\Delta-d\Delta_{\mathrm{c}}$}. 
\end{cases}
$$
It follows from these definitions that 
$$
E_{n}^{-}(z,0;d) 
= 
\begin{cases}
- \Delta_{\mathrm{c}}n + \varepsilon_{0}   
& \text{if $
\varepsilon_{1}-\varepsilon_{0}\ge
\Delta-d\Delta_{\mathrm{c}}$}, \\ 
- \Delta_{\mathrm{c}}n + \varepsilon_{1}-\Delta 
& \text{if $
\varepsilon_{1}-\varepsilon_{0}<
\Delta-d\Delta_{\mathrm{c}}$},  
\end{cases}
$$ 
and 
\begin{align*}
E_{n}^{+}(z,t;d) 
\ge& E_{n}^{+}(z,0;d) \\ 
=& 
\begin{cases}
- \Delta_{\mathrm{c}}n + \varepsilon_{1}-\Delta 
& \text{if $
\varepsilon_{1}-\varepsilon_{0}\ge
\Delta-d\Delta_{\mathrm{c}}$}, \\ 
- \Delta_{\mathrm{c}}n + \varepsilon_{0}   
& \text{if $
\varepsilon_{1}-\varepsilon_{0}<
\Delta-d\Delta_{\mathrm{c}}$}. 
\end{cases}
\end{align*} 
The later inequality means that the candidates 
of superradiant ground state energy are 
only $E_{n}^{-}(z,t;d)$s. 

We define two spatiotemporal 
domains $\mathbb{D}^{\mathrm{wc}}_{mn}(d)$ 
and $\mathbb{D}^{\mathrm{sc}}_{mn}(d)$ 
for non-negative integers $m$ and $n$ with $\max\{d,m\}
<n$ as 
\begin{align*}
\mathbb{D}^{\mathrm{wc}}_{mn}(d) := 
\left\{
(z,t)\, |\, \right.
&E_{m}^{0}(z,t;d) < E_{n}^{-}(z,t;d)\,\,\, 
\text{if $m<d$};\,\,\,  \\ 
&
E_{m}^{-}(z,t;d) < E_{n}^{-}(z,t;d)\,\,\, 
\text{if $m\ge d$}
\left.\right\}
\end{align*} 
and 
\begin{align*}
\mathbb{D}^{\mathrm{sc}}_{mn}(d) := 
\left\{
(z,t)\, |\, \right.
&E_{m}^{0}(z,t;d) > E_{n}^{-}(z,t;d)\,\,\, 
\text{if $m<d$};\,\,\, \\ 
&E_{m}^{-}(z,t;d) > E_{n}^{-}(z,t;d)\,\,\, 
\text{if $m\ge d$}
\left.\right\}
\end{align*} 
respectively.    

\subsection{In the case $d=1$}
\label{subsec:d=1}

In this subsection we investigate 
the behavior of the Dicke-type 
energy level crossings in the case $d=1$. 
To do that, we introduce some positive numbers 
and some domains of the space-time. 
Then, we divide the whole space of the space-time 
into three classes, namely, 
the weak coupling regime, 
the strong coupling regime, 
and the critical regime:

For each natural number $n$, 
we define a positive number 
$C_{0n}^{0}$ by 
$$
C_{0n}^{0}:=
\begin{cases}
\Delta_{\mathrm{c}}^{2}n-\Delta_{\mathrm{c}}
\left(\varepsilon_{1}-
\varepsilon_{0}+\Delta_{\mathrm{c}}
-\Delta\right) 
& \quad \\ 
\qquad\qquad\qquad\qquad\qquad
\text{if $\varepsilon_{1}-\varepsilon_{0}\ge 
\Delta-\Delta_{\mathrm{c}}$,} 
&\quad \\ 
\Delta_{\mathrm{c}}^{2}(n+1)-\Delta_{\mathrm{c}}
\left(\varepsilon_{1}-
\varepsilon_{0}+\Delta_{\mathrm{c}}
-\Delta\right) 
& \quad \\ 
\qquad\qquad\qquad\qquad\qquad
\text{if $\varepsilon_{1}-\varepsilon_{0}< 
\Delta-\Delta_{\mathrm{c}}$.} & \quad 
\end{cases}
$$
We define three domains 
$\mathcal{D}^{\mathrm{wc}}_{0n}(1)$, 
$\mathcal{D}^{\mathrm{sc}}_{0n}(1)$, 
and $\mathcal{D}^{0}_{0n}(1)$ 
of the space-time as follows: 
the spatiotemporal domain 
$\mathcal{D}_{0n}^{\mathrm{wc}}(1)$ 
for the weak coupling regime is given by
$\mathcal{D}^{\mathrm{wc}}_{0n}(1):= 
\left\{
(z,t) \, |\, 
|\Omega(z,t)|^{2} < C_{0n}^{0}
\right\}$, 
and the domain 
$\mathcal{D}_{0n}^{\mathrm{sc}}(1)$ 
for the strong coupling regime by 
$\mathcal{D}^{\mathrm{sc}}_{0n}(1):= 
\left\{
(z,t) \, |\, 
|\Omega(z,t)|^{2} > C_{0n}^{0}
\right\}$. 
The domain 
$\mathcal{D}_{0n}^{\mathrm{cr}}(1)$ 
for the critical regime is defined by 
$\mathcal{D}^{\mathrm{cr}}_{0n}(1):= 
\left\{
(z,t) \, |\, 
|\Omega(z,t)|^{2} = C_{0n}^{0}
\right\}$.

The following theorem says 
that how the Dicke-type energy level crossing 
takes place is completely determined: 
\textit{Let us suppose $1 < n$ now. Then, 
the spatiotemporal domain 
$\mathcal{D}^{\mathrm{wc}}_{0n}(1)$ 
is equal to the domain $\mathbb{D}^{\mathrm{wc}}_{0n}(1)$, 
and the spatiotemporal domain 
$\mathcal{D}^{\mathrm{sc}}_{0n}(1)$ 
to the domain 
$\mathbb{D}^{\mathrm{sc}}_{0n}(1)$, 
i.e., $\mathcal{D}^{\mathrm{wc}}_{0n}(1) = 
\mathbb{D}^{\mathrm{wc}}_{0n}(1)$ and 
$\mathcal{D}^{\mathrm{sc}}_{0n}(1) = 
\mathbb{D}^{\mathrm{sc}}_{0n}(1)$. 
Namely, the energy level crossing takes place as} 
\begin{align}
& E_{0}^{0}(z,t;1) < E_{n}^{-}(z,t;1)\,\,\, 
\textit{if and only if $(z,t)$ in 
$\mathcal{D}_{0n}^{\mathrm{wc}}(1)$},  
\label{eq:before-DC0} \\ 
& E_{0}^{0}(z,t;1) = E_{n}^{-}(z,t;1)\,\,\, 
\textit{if and only if $(z,t)$ in 
$\mathcal{D}_{0n}^{\mathrm{cr}}(1)$},   
\label{eq:0-DC0} \\ 
& E_{0}^{0}(z,t;1) > E_{n}^{-}(z,t;1)\,\,\, 
\textit{if and only if $(z,t)$ in 
$\mathcal{D}_{0n}^{\mathrm{sc}}(1)$}.  
\label{eq:after-DC0}
\end{align}
These inequalities 
(\ref{eq:before-DC0})--(\ref{eq:after-DC0}) 
guarantee the Dicke-type 
energy level crossing 
because $E_{0}^{0}(z,t;1)$ 
and $E_{n}^{-}(z,t;1)$ are continuous functions 
of the space-time point $(z,t)$.

When the above Dicke-type energy level crossing 
between $E_{0}^{0}(z,t;1)$ and $E_{n}^{-}(z,t;1)$ 
takes place, 
there is certainly an energy level crossing 
between $E_{m}^{-}(z,t;1)$ and $E_{n}^{-}(z,t;1)$ 
for a natural number $m$ with $1 < m < n$. 
To show it, we introduce two positive constants and 
define two domains of the space-time:

For natural numbers $m, n$ with $m < n$, 
we set positive constants $C_{mn}^{\mathrm{wc}}$ 
and $C_{mn}^{\mathrm{sc}}$ as 
\begin{align*}
C_{mn}^{\mathrm{wc}} 
= 
\begin{cases}
\Delta_{\mathrm{c}}^{2}
\left\{
{\displaystyle \frac{m+n}{2}}+
\sqrt{\displaystyle \left(\frac{m+n}{2}\right)^{2}
+\frac{K^{2}}{2\Delta_{\mathrm{c}}^{2}}}
\right\} 
& \quad \\ 
\qquad\qquad\qquad\qquad\qquad
\text{if $\varepsilon_{1}-\varepsilon_{0}\ge 
\Delta-\Delta_{\mathrm{c}}$,} 
&\quad \\ 
\Delta_{\mathrm{c}}^{2}
\left\{
{\displaystyle \frac{m+n+2}{2}}+
\sqrt{\displaystyle \left(\frac{m+n+2}{2}\right)^{2}
+\frac{K^{2}}{2\Delta_{\mathrm{c}}^{2}}}
\right\} 
& \quad \\ 
\qquad\qquad\qquad\qquad\qquad
\text{if $\varepsilon_{1}-\varepsilon_{0}< 
\Delta-\Delta_{\mathrm{c}}$,} 
&\quad 
\end{cases}
\end{align*} 
and 
\begin{align*}
C_{mn}^{\mathrm{sc}} 
= 
\begin{cases}
\Delta_{\mathrm{c}}^{2}
\left\{
m+n+2\sqrt{\displaystyle 
mn+
\frac{K^{2}}{
4\Delta_{\mathrm{c}}^{2}}}
\right\} 
& \quad \\ 
\qquad\qquad\qquad\qquad\qquad
\text{if $\varepsilon_{1}-\varepsilon_{0}\ge 
\Delta-\Delta_{\mathrm{c}}$,} 
&\quad \\ 
\Delta_{\mathrm{c}}^{2}
\left\{
m+n+2+2\sqrt{\displaystyle 
m+n+mn
+\frac{K^{2}}{
4\Delta_{\mathrm{c}}^{2}}}
\right\} 
& \quad \\ 
\qquad\qquad\qquad\qquad\qquad
\text{if $\varepsilon_{1}-\varepsilon_{0}< 
\Delta-\Delta_{\mathrm{c}}$,} 
&\quad 
\end{cases}
\end{align*}
where $K=\varepsilon_{1}-\varepsilon_{0}
+\Delta_{\mathrm{c}}-\Delta$. 
We give two spatiotemporal domains 
$\mathcal{D}_{mn}^{\mathrm{wc}}(1)$ and 
$\mathcal{D}_{mn}^{\mathrm{sc}}(1)$ 
for non-negative integers $m, n$ 
with $m < n$ in the following: 
The domain $\mathcal{D}_{mn}^{\mathrm{wc}}(1)$ 
for the weak coupling regime is given by 
$\mathcal{D}_{mn}^{\mathrm{wc}}(1) 
= 
\left\{
(z,t) \, |\, 
0 \le 
|\Omega(z,t)|^{2} < C_{mn}^{\mathrm{wc}}
\right\}$, 
and the domain $\mathcal{D}_{mn}^{\mathrm{sc}}(1)$ 
for the strong coupling regime by 
$\mathcal{D}_{mn}^{\mathrm{sc}}(1) 
= 
\left\{
(z,t) \, |\, 
|\Omega(z,t)|^{2} > C_{mn}^{\mathrm{sc}}
\right\}$.  
Then, as far as such $m$ goes, 
the following theorem gives 
a sufficient condition so that 
the crossing between $E_{m}^{-}(z,t;1)$ 
and $E_{n}^{-}(z,t;1)$ occurs: 
\textit{Let $1 < m < n$ now. 
Then, the spatiotemporal domain 
$\mathcal{D}^{\mathrm{wc}}_{mn}(1)$ 
is included in the domain 
$\mathbb{D}^{\mathrm{wc}}_{mn}(1)$, 
and the domain $\mathcal{D}^{\mathrm{sc}}_{mn}(1)$ 
in the domain 
$\mathbb{D}^{\mathrm{sc}}_{mn}(1)$, 
i.e., $\mathcal{D}^{\mathrm{wc}}_{mn}(1) 
\subset 
\mathbb{D}^{\mathrm{wc}}_{mn}(1)$ 
and 
$\mathcal{D}^{\mathrm{sc}}_{mn}(1) \subset 
\mathbb{D}^{\mathrm{sc}}_{mn}(1)$}: 
\begin{align}
& E_{m}^{-}(z,t;1) < E_{n}^{-}(z,t;1)\,\,\, 
\textit{for $(z,t)$ in $\mathcal{D}_{mn}^{\mathrm{wc}}(1)$},  
\label{eq:before-DC} \\ 
& E_{m}^{-}(z,t;1) > E_{n}^{-}(z,t;1)\,\,\, 
\textit{for $(z,t)$ in $\mathcal{D}_{mn}^{\mathrm{sc}}(1)$}. 
\label{eq:after-DC}
\end{align}

\begin{figure}[htbp]
  \begin{center}
  \resizebox{70mm}{!}{\includegraphics{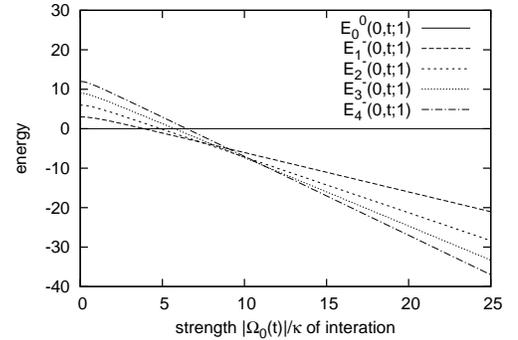}}
  \end{center}
  \caption{Dicke-type energy level crossings 
for $\gamma(z)$$=$$\cos 2\pi z$. 
Energies $E_{0}^{0}(0,t;1)$ (solid line), 
$E_{1}^{-}(0,t;1)$ (dashed), 
$E_{2}^{-}(0,t;1)$ (short-dashed), 
$E_{3}^{-}(0,t;1)$ (dotted),   
and $E_{4}^{-}(0,t;1)$ (dashed-dotted). 
The physical parameters are set as 
$\varepsilon_{0}=0$, $\varepsilon_{1}=6\kappa$, 
$\Delta=1\kappa$, $\Delta_{\mathrm{c}}=-3\kappa$ 
with a unit $\kappa$. 
The position $z$ is fixed at $z=0$.}
  \label{fig:Es-a} 
\end{figure}

\begin{figure}[htbp]
  \begin{center}
  \resizebox{70mm}{!}{\includegraphics{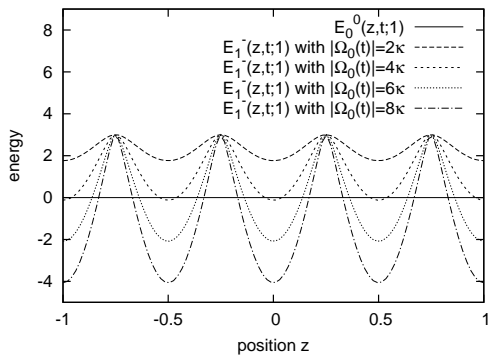}}
  \end{center}
  \caption{Dicke-type energy level crossings 
for $\gamma(z)=\cos 2\pi z$. 
Energies $E_{0}^{0}(z,t;1)$ (solid line), 
$E_{1}^{-}(0,t;1)$ with $|\Omega_{0}(t)|
=2\kappa$ (dashed), 
$E_{1}^{-}(0,t;1)$ with $|\Omega_{0}(t)|
=4\kappa$ (short-dashed), 
$E_{1}^{-}(0,t;1)$ with $|\Omega_{0}(t)|
=6\kappa$ (dptteed), 
and $E_{1}^{-}(0,t;1)$ with $|\Omega_{0}(t)|
=8\kappa$ (dashed-dotted). 
The time is fixed so that 
$|\Omega_{0}(t)|=2\kappa, 4\kappa, 6\kappa, 8\kappa$.}
  \label{fig:Es-b} 
\end{figure}

These two theorems will be proved at the end of 
this subsection. 
But, before proving them, 
we concretely see some physical situation that they 
tell us. 
We employ $\cos2\pi z$ as $\gamma(z)$. 
For the position fixed at $z=0$, 
the five energies 
$E_{0}^{0}(0,t;1)$ and $E_{n}^{-}(0,t;1)$, 
$n=1, 2, 3, 4$, are numerically calculated as in 
Fig.\ref{fig:Es-a}. 
In the case where the strength $|\Omega_{0}(t)|$ 
is given by 
$|\Omega_{0}(t)| = 2\kappa, 4\kappa, 
6\kappa, 8\kappa$, 
the two energies 
$E_{0}^{0}(z,t;1)$ and $E_{1}^{-}(z,t;1)$ are 
in Fig.\ref{fig:Es-b}. 

Suppose that $\varepsilon_{1}-\varepsilon_{0}
\ge \Delta - \Delta_{\mathrm{c}}$ now. 
Let our atom be in the state with the energy 
$E_{m}^{-}(z_{0},t_{0};1)$ 
at an initial space-time point $(z_{0},t_{0})$ 
in the domain $\mathcal{D}_{mn}^{\mathrm{wc}}(1)$ 
for a non-negative integer $m$ 
and a positive integer $n$ with $0\le m<n$. 
The atom is apt to sit in the state 
with energy $E_{m}^{-}(z_{0},t_{0};1)$ 
because being in the state 
with energy $E_{m}^{-}(z_{0},t_{0};1)$ is more stable 
than in the state 
with energy $E_{n}^{-}(z_{0},t_{0};1)$. 
Once, however, a space-time point $(z,t)$ 
plunges into the domain 
$\mathcal{D}_{mn}^{\mathrm{sc}}(1)$, 
the energy level crossing takes place 
as shown in its process 
(\ref{eq:before-DC0})--(\ref{eq:after-DC0}). 
Thus, being in the state with energy $E_{m}^{-}(z,t;1)$ 
is \textit{not} stable any longer, 
and thus, it goes down to the state with energy 
$E_{n}^{-}(z,t;1)$. 
This descent can be caused by cavity decay 
even if we cannot expect atomic decay as 
pointed out  in Ref.\onlinecite{H-R}.    
The circumstance for such a cooling is usually 
not in thermal equilibrium, and thus, 
the temperature which this system loses is not 
given by thermodynamic temperature. 
Nevertheless, according to the thermodynamics low 
as in Eq.(5.1) of Ref.\onlinecite{MvdS} 
and Eq.(2.1) of Ref.\onlinecite{B-C-T}, 
we roughly estimate the effective temperature 
$\Delta T_{m\to n}$ coming from the descent. 
Then, we can expect that 
the crossing between 
$E_{m}^{-}(z,t;1)$ and $E_{n}^{-}(z,t;1)$ makes 
the temperature go down at most $\Delta T_{m\to n}$ 
estimated at: 
\begin{align}
\Delta T_{m\to n} 
\approx&
\frac{2(n-m)}{k_{\mathrm{B}}}
\Biggl|
\frac{|\Omega(z,t)|^{2}}{
\Upsilon_{m}(z_{0},t_{0};1)+\Upsilon_{n}(z,t;1)} 
- |\Delta_{\mathrm{c}}|
\Biggr| \notag \\ 
&+ 
\frac{2m}{k_{\mathrm{B}}}
\Biggl|
\frac{|\Omega(z,t)|^{2}
-|\Omega(z_{0},t_{0})|^{2}}{
\Upsilon_{m}(z_{0},t_{0};1)+\Upsilon_{n}(z,t;1)}
\Biggr|
\label{eq:temperature-loss-mn} 
\end{align}
for the point $(z_{0},t_{0})$ in 
the spatiotemporal domain 
$\mathcal{D}_{mn}^{\mathrm{wc}}(1)$ and 
the point $(z,t)$ in the domain 
$\mathcal{D}_{mn}^{\mathrm{sc}}(1)$, 
of course, 
provided that there is nothing to obstruct the temperature 
loss. 
Here $k_{\mathrm{B}}$ is the Boltzmann constant 
and $\Upsilon_{\ell}(z,t;1)=(1/2)
\sqrt{K^{2}+4|\Omega(z,t)|^{2}\ell\,\,}$ 
the generalized Rabi frequency 
(\ref{eq:GRF}) for each natural 
number $\ell$.

\begin{figure}[htbp]
  \begin{center}
  \resizebox{70mm}{!}{\includegraphics{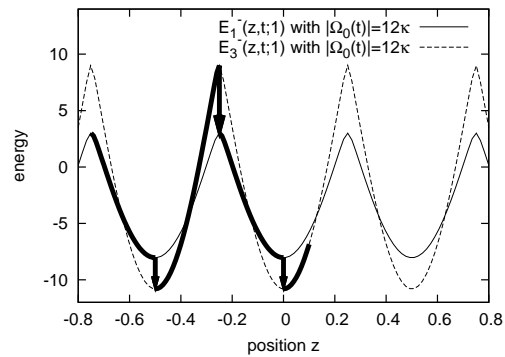}}
  \end{center}
  \caption{Energy decay between $E_{1}^{-}(z,t;1)$ 
(solid line) and $E_{3}^{-}(z,t;1)$ (dashed). 
The physical parameters are set as 
in Fig.\ref{fig:Es-a} and the strength 
$|\Omega_{0}(t)|$ is fixed at $12\kappa$. 
$\gamma(z)=\cos 2\pi z$ in $\Omega(z,t)$. 
The thick arrows stand for 
the temperature loss. 
}
  \label{fig:superradiant-mn-pra} 
\end{figure}

Conversely, let our atom be in the state with the energy 
$E_{n}^{-}(z_{0},t_{0};1)$ at an initial space-time point 
$(z_{0},t_{0})$ in the domain 
$\mathcal{D}_{mn}^{\mathrm{sc}}(1)$. 
Then, since a process reverse to the above energy level 
crossing takes place if the coupling regime recoils 
from the spatiotemporal domain 
$\mathcal{D}_{mn}^{\mathrm{sc}}(1)$ 
to the domain $\mathcal{D}_{mn}^{\mathrm{wc}}(1)$, 
this reverse crossing can also carry away 
the temperature $\Delta T_{n\to m}$ 
of the atom so that $\Delta T_{n\to m}=
\Delta T_{m\to n}$
for the point $(z_{0},t_{0})$ in 
the spatiotemporal domain 
$\mathcal{D}_{mn}^{\mathrm{sc}}(1)$ 
and the point $(z,t)$ in 
the domain 
$\mathcal{D}_{mn}^{\mathrm{wc}}(1)$. 
We illustrate this situation for 
$E_{1}^{-}(z,t;1)$ and $E_{3}^{-}(z,t;1)$ 
in Fig.\ref{fig:superradiant-mn-pra}. 
The diagrammatic illustration about temperature loss 
is represented by the thick arrows 
in Fig.\ref{fig:superradiant-mn-pra}.
Therefore, our arguments say that there is a possibility 
of the following mechanism for superradiant cooling:  
Let a space-time point $(z_{2\ell+1},t_{2\ell+1})$ 
be in the domain 
$\mathcal{D}_{01}^{\mathrm{wc}}(1)$ 
and a space-time point $(z_{2\ell+2},t_{2\ell+2})$ 
in the domain 
$\mathcal{D}_{01}^{\mathrm{sc}}(1)$, 
respectively, for each $\ell=0, 1, \cdots, N-1$ with 
a natural number $N$. 
The Dicke-type energy level crossings and 
their reverse crossings may carry away 
the temperature of the atom:    
\begin{equation}
\frac{2}{k_{\mathrm{B}}}
\sum_{\nu=1}^{2N}
\frac{|\Omega(z_{\nu},t_{\nu})|^{2}}{|K|/2
+\Upsilon_{1}(z_{\nu},t_{\nu};1)} 
- \frac{4N}{k_{\mathrm{B}}}|\Delta_{\mathrm{c}}|
\label{eq:superradiant-cooling}
\end{equation}
at most by Eq.(\ref{eq:temperature-loss-mn}) 
if $K\ge 0$. 
We can make similar argument 
when $K<0$.   

\begin{figure}[htbp]
  \begin{center}
    \resizebox{70mm}{!}{\includegraphics{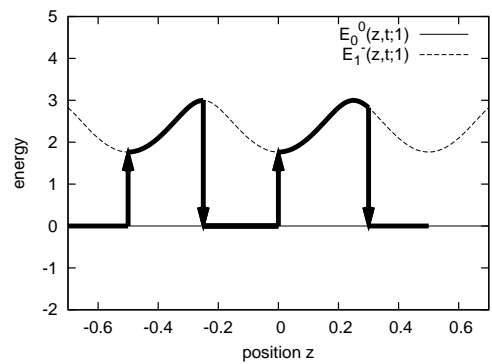}}
  \end{center}
  \caption{Cavity-induced atom cooling 
(bold line) in $\mathbb{D}_{01}^{\mathrm{wc}}(1)$. 
Energies $E_{0}^{0}(t,z;1)$ (solid line) 
and $E_{0}^{-}(z,t;1)$ (dashed). 
The physical parameters are set as 
in Fig.\ref{fig:Es-a}.}
  \label{fig:ritsch-pra}
\end{figure}

\begin{figure}[htbp]
  \begin{center}
  \resizebox{70mm}{!}{\includegraphics{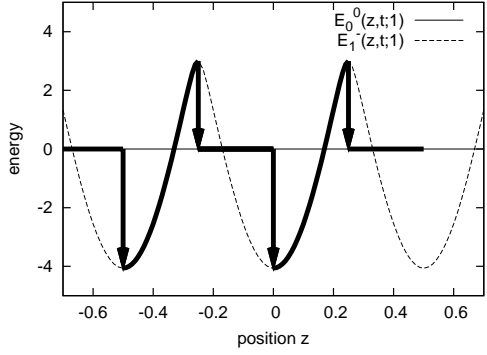}}
  \end{center}
  \caption{Dicke-type crossings and 
cavity-induced atom cooling (bold line). 
Energies $E_{0}^{0}(z,t;1)$ (solid line) 
and $E_{0}^{-}(z,t;1)$ (dashed). 
The physical parameters are set as 
in Fig.\ref{fig:Es-a}.}
  \label{fig:superradiant-pra}
\end{figure}

\begin{figure}[htbp]
    \begin{center}
      \resizebox{85mm}{!}{\includegraphics{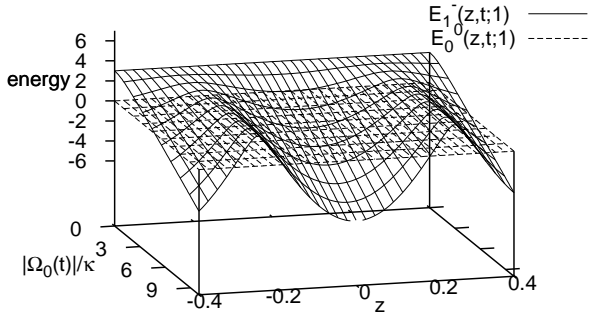}}
      \caption{Surfaces of energies 
$E_{0}^{0}(z,t;1)=0$ (dashed line) and 
$E_{1}^{-}(z,t;1)$ (solid). 
The physical parameters are set as 
in Fig.\ref{fig:Es-a}.}
      \label{fig:surface1}
    \end{center}
\end{figure}

\begin{figure}[htbp]
    \begin{center}
      \resizebox{85mm}{!}{\includegraphics{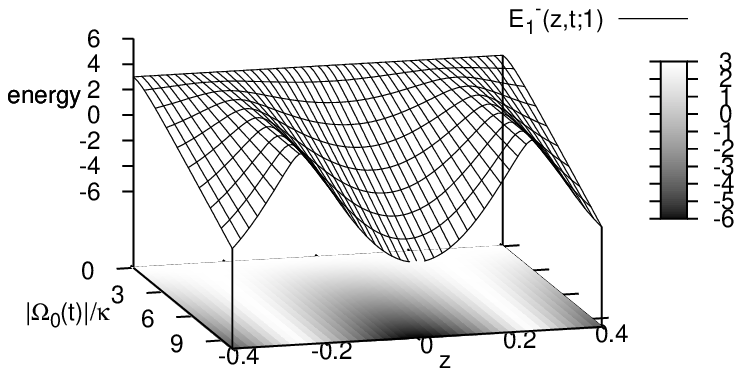}}
      \caption{Surface of energy $E_{1}^{-}(z,t;1)$ 
(solid line), where $E_{0}^{0}(z,t;1)=0$. 
The physical parameters are set as 
in Fig.\ref{fig:Es-a}.}
      \label{fig:surface2}
   \end{center}
\end{figure}

\begin{figure}[htbp]
    \begin{center}
      \resizebox{85mm}{!}{\includegraphics{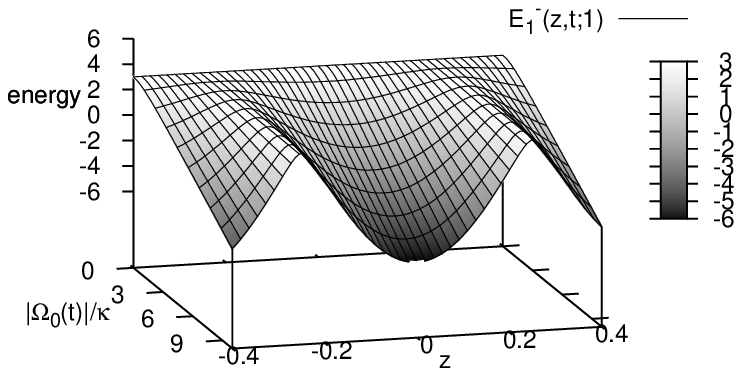}}
      \caption{Surface of energy $E_{1}^{-}(z,t;1)$ 
(solid line), where $E_{0}^{0}(z,t;1)=0$. 
The physical parameters are set as 
in Fig.\ref{fig:Es-a}.}
      \label{fig:surface3}
   \end{center}
\end{figure}

When the whole space of the space-time 
is $\mathbb{D}_{01}^{\mathrm{wc}}(1)$, 
we find the Sisyphus-type mechanism 
in the energy spectrum as the group of Ritsch 
pointed out in Ref.\onlinecite{H-R}. 
To see the diagrammatic representation 
we consider a concrete example now. 
For the strength $|\Omega_{0}(t)|=2\kappa$ 
we have the cavity-induced atom cooling as in 
Fig.\ref{fig:ritsch-pra}. 
Up-arrows in Fig.\ref{fig:ritsch-pra} 
represent the energy that the atom coupled with 
photons gains by photon absorption, namely, we have to 
throw a driving laser to the atom for its excitement. 
Down-arrows mean the energy loss caused by 
cavity decay. 
In this domain $\mathbb{D}_{01}^{\mathrm{wc}}(1)$ 
of the space-time 
we cannot find such a sequence 
$\left\{(z_{\nu},t_{\nu})\right\}_{\nu=1}^{2L}$ 
of the space-time points. 
Thus, Eq.(\ref{eq:superradiant-cooling}) 
does not work.  
As shown in the following concrete example, 
on the other hand, 
we can expect the sequence 
$\left\{(z_{\nu},t_{\nu})\right\}_{\nu=1}^{2N}$:
If we take a strength as $|\Omega_{0}|=8\kappa$, 
then Fig.\ref{fig:ritsch-pra} changes to 
Fig.\ref{fig:superradiant-pra}, and then, 
we can find the sequence 
$\left\{(z_{\nu},t_{\nu})\right\}_{\nu=1}^{2N}$ 
in Fig.\ref{fig:superradiant-pra}. 
Compare Fig.\ref{fig:ritsch-pra} 
and Fig.\ref{fig:superradiant-pra}. 
In Fig.\ref{fig:superradiant-pra} 
we can make only down-arrows without any uparrow. 
It means that the system loses energy only 
because of cavity decay. 
Namely, we do not have to throw the driving laser 
to the atom in the cavity for its excitement. 
We note this type emission comes from a kind of 
superradiance \cite{PL,hir01,hir02}. 
Thus, this energy-spectral property 
may give a mechanism for another superradiant 
cooling for a two-level atom 
in the strong coupling regime, 
as well as self-organized, cooperative atoms 
 \cite{DR,DR02}. 
As we can realize it from Figs.\ref{fig:Es-b} and 
\ref{fig:superradiant-pra}, 
the energy loss by cavity decay gets large as the strength 
$|\Omega_{0}(t)|$ of the coupling grows large. 
We give the surfaces of $E_{0}^{0}(z,t;1)$ and 
$E_{1}^{-}(z,t;1)$ as functions of $(z,t)$ 
in Figs.\ref{fig:surface1}, \ref{fig:surface2}, 
and \ref{fig:surface3}.   

We prove our two theorems on the Dicke-type energy level 
crossing now: 
To make the calculations below simple, 
we set $K$ as $K:=\varepsilon_{1}-\varepsilon_{0}
+\Delta_{\mathrm{c}}-\Delta$ again. 

We give a proof of crossings 
(\ref{eq:before-DC0})--(\ref{eq:after-DC0}) first. 
Let us suppose $K\ge 0$ now. 
It is easy to show the equation, 
\begin{equation}
E_{0}^{0}(z,t;1)-E_{n}^{-}(z,t;1) 
= \Delta_{\mathrm{c}}n -\frac{1}{2}K
+\Upsilon_{n}(z,t;1),  
\label{eq:E_0-E_1-0}
\end{equation}
where $\Upsilon_{n}(z,t;1)$ is the 
generalized Rabi frequency (\ref{eq:GRF}). 
Let the symbol $\sharp$ denote either 
$>$, $=$, or $<$. 
After multiplying both sides of the expression 
$|\Omega(z,t)|^{2}\,\,\sharp\,\,\Delta_{\mathrm{c}}^{2}n
-\Delta_{\mathrm{c}}K$ by $4n$, 
add the term $K^{2}$ to both sides of 
the multiplied expression. 
Then, we know that the expression 
$|\Omega(z,t)|^{2}\,\,\sharp\,\,\Delta_{\mathrm{c}}^{2}n
-\Delta_{\mathrm{c}}K$, is equivalent to the expression 
$K^{2}+4n\Omega(z,t)^{2}\,\,\sharp\,\, 4
(\Delta_{\mathrm{c}}^{2}n^{2}
-\Delta_{\mathrm{c}}Kn+K^{2}/4)$. 
Since $-\Delta_{\mathrm{c}}n+K/2 \ge 0$, 
we can take the square root of the last expression, 
and thus, it is equivalent to 
$2\Upsilon_{n}(z,t;1)\,\,
\sharp\,\,2(-\Delta_{\mathrm{c}}n+K/2)$. 
Therefore, Eq.(\ref{eq:E_0-E_1-0}) says that 
\begin{equation}
E_{0}^{0}(z,t;1)-E_{n}^{-}(z,t;1)\,\,\sharp\,\, 0\,\,\, 
\Longleftrightarrow\,\,\,
|\Omega(z,t)|^{2}\,\,\sharp\,\, 
\Delta_{\mathrm{c}}^{2}n-\Delta_{\mathrm{c}}K, 
\label{eq:081207-1}
\end{equation}
where ``RHS''$\Longleftrightarrow$``LHS'' means that 
the right hand side of the equation 
\textit{is equivalent to} 
the left hand side. 
The equivalence (\ref{eq:081207-1}) 
brings the crossings 
(\ref{eq:before-DC0})--(\ref{eq:after-DC0}). 

Let us suppose $K<0$ now. 
Then, in the same way we did in the case $K>0$, 
we have the equivalence: 
\begin{align}
& E_{0}^{0}(z,t;1) - E_{n}^{-}(z,t;1)\,\,\sharp\,\, 0 \notag \\ 
\Longleftrightarrow 
&\Upsilon_{n+1}(z,t;1)\,\,\sharp\,\, 
-\Delta_{\mathrm{c}}(n+1)+\frac{1}{2}K.
\label{eq:081025-1}
\end{align}
The term $-\Delta_{\mathrm{c}}(n+1)+K/2$ 
is positive because we assumed the condition 
(\ref{eq:assumption}) and $\Delta_{\mathrm{c}}<0$. 
Thus, by taking the square of both sides of 
the right expression of the equivalence 
(\ref{eq:081025-1}) and following the way 
we did in the case $K\ge 0$, 
we know the expression 
$E_{0}^{0}(z,t;1) - E_{n}^{-}(z,t;1)\,\,\sharp\,\, 0 $ 
is equivalent to the expression 
$|\Omega(z,t)|^{2}\,\,\sharp\,\, \Delta_{\mathrm{c}}^{2}(n+1)
-\Delta_{\mathrm{c}}K$, which implies our desired result.

We consider the proofs of the crossing 
(\ref{eq:before-DC}) and (\ref{eq:after-DC}) next. 
Let us suppose $d \le m < n$. 
A direct calculation leads to 
\begin{align}
&E_{m}^{-}(z,t;1)-E_{n}^{-}(z,t;1) \notag \\ 
=& \Delta_{\mathrm{c}}(n-m) 
+\Upsilon_{n}(z,t;1)
-\Upsilon_{m}(z,t;1) \notag \\ 
=& 
(n-m)
\left[
\Delta_{\mathrm{c}} + 
\frac{|\Omega(z,t)|^{2}}{
\Upsilon_{n}(z,t;1)
+\Upsilon_{m}(z,t;1)
}
\right].
\label{eq:081014-0}
\end{align}
Since $n>m$ and 
$|\Delta_{\mathrm{c}}|=-\Delta_{\mathrm{c}}$, 
Eq.(\ref{eq:081014-0}) says that 
the expression $E_{m}^{-}(z,t;1)-E_{n}^{-}(z,t;1)\,\,
\sharp\,\, 0$ is equivalent to the expression 
$|\Omega(z,t)|^{2}
\left(\Upsilon_{n}(z,t;1)+\Upsilon_{m}(z,t;1)
\right)^{-1}\,\,\sharp\,\,|\Delta_{\mathrm{c}}|$. 
Multiplying both sides of this by 
$2|\Delta_{\mathrm{c}}|^{-1}
|\Omega(z,t)|^{-1}
(\Upsilon_{n}(z,t;1)
+\Upsilon_{m}(z,t;1))$,
we realize that the later expression is equivalent 
to the expression  
$2|\Omega(z,t)|^{-1}/|\Delta_{\mathrm{c}}|$
$\sharp$ 
$\sqrt{(K/|\Omega(z,t)|)^{2}+4n}+
\sqrt{(K/|\Omega(z,t)|)^{2}+4m}$. 
Hence it follows from these equivalences that 
\begin{widetext}
\begin{equation}
E_{m}^{-}(z,t;1)-E_{n}^{-}(z,t;1)\,\,\sharp\,\, 0\,\,\, 
\Longleftrightarrow\,\,\, 
2\frac{|\Omega(z,t)|}{|\Delta_{\mathrm{c}}|} 
- \sqrt{\left(\frac{K}{|\Omega(z,t)|}\right)^{2}
+4m\,}
\,\,\,\,\,\sharp\,\,\,\,\, 
\sqrt{\left(\frac{K}{|\Omega(z,t)|}\right)^{2}
+4n\,}. 
\label{eq:081014-1}
\end{equation}
\end{widetext}

Suppose $K \ge 0$ now. 
Here we note a simple inequality 
for non-negative numbers $A, B$, 
and $C$: 
$\left(\sqrt{A+B\,}+\sqrt{A+C\,}\right)^{2}
\ge A+B+A+C$.  
Using this, 
we can show that the expression 
$4\left(|\Omega(z,t)|/|\Delta_{\mathrm{c}}\right)^{2} 
< \left(K/|\Omega(z,t)|\right)^{2}+4n+
\left(K/|\Omega(z,t)|\right)^{2}+4m$ implies 
the expression 
$4\left(|\Omega(z,t)|/|\Delta_{\mathrm{c}}\right)^{2} 
< \{\sqrt{(K/|\Omega(z,t)|)^{2}+4n\,}+
\sqrt{(K/|\Omega(z,t)|)^{2}+4m\,}\}^{2}$.  
Taking the square root of 
both sides of this inequality 
implies the expression, 
$2(|\Omega(z,t)|/|\Delta_{\mathrm{c}}|)<
\sqrt{(K/|\Omega(z,t)|)^{2}+4n\,}+
\sqrt{(K/|\Omega(z,t)|)^{2}+4m\,}$. 
Therefore, we obtain the following: 
\begin{widetext}
\begin{align}
& 4\left(\frac{|\Omega(z,t)|}{
|\Delta_{\mathrm{c}}|}\right)^{2} 
<  
\left(\frac{K}{
|\Omega(z,t)|}\right)^{2}+4n
+\left(\frac{K}{
|\Omega(z,t)|}\right)^{2}+4m  \notag \\ 
\Longrightarrow\,\,\,  & 
2\frac{|\Omega(z,t)|}{
|\Delta_{\mathrm{c}}|} 
- \sqrt{\left(\frac{K}{
|\Omega(z,t)|}\right)^{2}+4m\,}  
<  
\sqrt{\left(\frac{K}{
|\Omega(z,t)|}\right)^{2}+4n\,}, 
\label{eq:081014-2}
\end{align}
\end{widetext}
where ``RHS''$\Longrightarrow$``LHS'' means that 
the right hand side of the equation 
\textit{implies} the left hand side. 

We define a polynomial $g_{\mathrm{wc}}(r)$ by 
$g_{\mathrm{wc}}(r):=2r^{2}-2(m+n)r-
K^{2}/|\Delta_{\mathrm{c}}|^{2}$. 
Multiplying both sides of this by $2$, 
the inequality $g_{\mathrm{wc}}\left(
\left(|\Omega(z,t)|/
|\Delta_{\mathrm{c}}|\right)^{2}\right)<0$ 
is equivalent to the inequality 
$4(|\Omega(z,t)|/|\Delta_{\mathrm{c}}|)^{4}
<4(m+n)(|\Omega(z,t)|/|\Delta_{\mathrm{c}}|)^{2}
+2(K/|\Delta_{\mathrm{c}}|)^{2}$. 
Multiplying both sides of this newly obtained 
inequality by $(|\Delta_{\mathrm{c}}|/|\Omega(z,t)|)^{2}$ 
we reach the following: 
\begin{widetext}
\begin{align}
& g_{\mathrm{wc}}\left(
\left(\frac{|\Omega(z,t)|}{
|\Delta_{\mathrm{c}}|}\right)^{2}\right) 
< 0 \notag \\ 
\Longleftrightarrow\,\,\, 
& \quad 4\left(
\frac{|\Omega(z,t)|}{|\Delta_{\mathrm{c}}|}
\right)^{2} 
<
\left(
\frac{K}{|\Omega(z,t)|}
\right)^{2}
+4m 
+\left(
\frac{K}{|\Omega(z,t)|}
\right)^{2}
+4n. 
\label{eq:081014-3}
\end{align}
\end{widetext}

It follows from the implication (\ref{eq:081014-2}), 
and equivalences (\ref{eq:081014-1}) and 
(\ref{eq:081014-3}) that 
\begin{align}
& g_{\mathrm{wc}}\left(
\left(\frac{|\Omega(z,t)|}{
|\Delta_{\mathrm{c}}|}\right)^{2}\right) 
< 0 \notag \\ 
\Longrightarrow\,\,\, 
& E_{m}^{-}(z,t;1)-E_{n}^{-}(z,t;1)<0.
\label{eq:081014-4}
\end{align}
 
Since the point $r_{0}^{\mathrm{wc}}$ 
defined by $r_{0}^{\mathrm{wc}}:=\left(m+n\right)/2
+\sqrt{\left(\left(m+n\right)/2\right)^{2}
+K^{2}/\left(2\Delta_{\mathrm{c}}^{2}\right)}$ 
satisfies $g_{\mathrm{wc}}(r_{0}^{\mathrm{wc}})=0$, 
we have the inequality $g_{\mathrm{wc}}(r)
<g_{\mathrm{wc}}(r_{0}^{\mathrm{wc}})=0$ 
provided that $0\le r <r_{0}^{\mathrm{wc}}$.
This fact tells us that 
\begin{align}
& \left(\frac{|\Omega(z,t)|}{
|\Delta_{\mathrm{c}}|}\right)^{2}
<\frac{m+n}{2} 
+ \sqrt{
\left(\frac{m+n}{2}\right)^{2}
+\frac{K^{2}}{2\Delta_{\mathrm{c}}^{2}}\,} \notag \\ 
\Longrightarrow\,\,\, 
& 
g_{\mathrm{wc}}\left(
\left(\frac{|\Omega(z,t)|}{
|\Delta_{\mathrm{c}}|}\right)^{2}\right) 
< 0.
\label{eq:081014-5}
\end{align}
Therefore, we can conclude the inequality 
(\ref{eq:before-DC}) from implications 
(\ref{eq:081014-4}) and (\ref{eq:081014-5}). 

We prove the inequality (\ref{eq:after-DC}) next. 
Multiply both sides of the inequality 
$(|\Omega(z,t)|/|\Delta_{\mathrm{c}}|)^{2}
-(n-m)>(|\Omega(z,t)|/|\Delta_{\mathrm{c}}|)
\sqrt{(K/|\Omega(z,t)|)^{2}+4m\,}$ by $4$, 
and add $(K/|\Omega(z,t)|)^{2}$ to both sides 
of the multiplied inequality. 
Then, we know that the inequality 
$\left(|\Omega(z,t)|/|\Delta_{\mathrm{c}}|\right)^{2}
-(n-m)>\left(|\Omega(z,t)|/|\Delta_{\mathrm{c}}|\right)
\sqrt{(K/|\Omega(z,t)|)^{2}+4m\,}$ 
implies the inequality 
$4(|\Omega(z,t)|/|\Delta_{\mathrm{c}}|)^{2}
+(K/|\Omega(z,t)|)^{2}-4(n-m)
>(K/|\Omega(z,t)|)^{2}
+4(|\Omega(z,t)|/|\Delta_{\mathrm{c}}|)
\sqrt{(K/|\Omega(z,t)|)^{2}+4m\,}$, 
which is equivalent to the inequality 
\begin{align*}
& 
4\left(\frac{|\Omega(z,t)|}{|\Delta_{\mathrm{c}}|}\right)^{2}
-
4\left(\frac{|\Omega(z,t)|}{|\Delta_{\mathrm{c}}|}\right)
\sqrt{\left(\frac{K}{|\Omega(z,t)|}\right)^{2}+4m\,} \\ 
&\qquad 
+\left\{\left(\frac{K}{|\Omega(z,t)|}\right)^{2}
+4m\right\} \\ 
>& \left(\frac{K}{|\Omega(z,t)|}\right)^{2}
+4n.
\end{align*} 
Taking the square root of both sides of 
the last inequality, we reach the implication: 
\begin{widetext}
\begin{align}
& 
\left(\frac{|\Omega(z,t)|}{
|\Delta_{\mathrm{c}}|}\right)^{2}
-(n-m)  
>
\left(\frac{|\Omega(z,t)|}{
|\Delta_{\mathrm{c}}|}\right)
\sqrt{\left(\frac{K}{
|\Omega(z,t)|}\right)^{2}+4m\,} \notag \\ 
\Longrightarrow\,\,\, 
& 2\frac{|\Omega(z,t)|}{
|\Delta_{\mathrm{c}}|}
- \sqrt{\left(\frac{K}{
|\Omega(z,t)|}\right)^{2}+4m\,}  
> \sqrt{\left(\frac{K}{
|\Omega(z,t)|}\right)^{2}+4n\,}.
\label{eq:081014-2'}
\end{align}
\end{widetext}

We define a polynomial $g_{\mathrm{sc}}(r)$ by 
$g_{\mathrm{sc}}(r):=r^{2}-2(m+n)r+(n-m)^{2}-
K^{2}/|\Delta_{\mathrm{c}}|^{2}$ this time. 
We note the inequality $g_{\mathrm{sc}}\left(
\left(|\Omega(z,t)|/
|\Delta_{\mathrm{c}}|\right)^{2}\right)>0$ 
is equivalent to the inequality 
$(|\Omega(z,t)|/|\Delta_{\mathrm{c}}|)^{4}
-2(n-m)(|\Omega(z,t)|/|\Delta_{\mathrm{c}}|)^{2}
+(n-m)^{2}
>(|\Omega(z,t)|/|\Delta_{\mathrm{c}}|)^{2}
\left\{
(K/|\Omega(z,t)|)^{2}+4m
\right\}$. 
Here we added $4m(|\Omega(z,t)|/
|\Delta_{\mathrm{c}}|)^{2}
+\left(K/|\Delta_{\mathrm{c}}|\right)^{2}$ 
to both sides of 
$g_{\mathrm{sc}}\left(
\left(|\Omega(z,t)|/|\Delta_{\mathrm{c}}|
\right)^{2}\right)>0$, and then, 
the right hand side was factorized 
with the factor 
$(|\Omega(z,t)|/|\Delta_{\mathrm{c}}|)^{2}$. 
Thus, taking the square root of both side of this inequality, 
we obtain the following implication:
\begin{widetext}
\begin{align}
& g_{\mathrm{sc}}\left(
\left(\frac{|\Omega(z,t)|}{
|\Delta_{\mathrm{c}}|}\right)^{2}\right)
> 0 \notag \\ 
\Longrightarrow\,\,\, 
& \left(
\frac{|\Omega(z,t)|}{|\Delta_{\mathrm{c}}|}
\right)^{2}
-(n-m)  
>
\frac{|\Omega(z,t)|}{|\Delta_{\mathrm{c}}|}
\sqrt{
\left(\frac{K}{
|\Delta_{\mathrm{c}}|}\right)^{2}+4m\,}. 
\label{eq:081014-3'}
\end{align}
\end{widetext}

It follows from the implications (\ref{eq:081014-2'}) 
and (\ref{eq:081014-3'}), and equivalence 
(\ref{eq:081014-1}) that 
\begin{align}
& g_{\mathrm{sc}}\left(
\left(\frac{|\Omega(z,t)|}{
|\Delta_{\mathrm{c}}|}\right)^{2}\right)
> 0  \notag \\ 
\Longrightarrow\,\,\, 
& E_{m}^{-}(z,t;1)-E_{n}^{-}(z,t;1)>0.
\label{eq:081014-4'}
\end{align}
 
Since $r_{0}^{\mathrm{sc}}:=m+n
+\sqrt{4mn+K^{2}/\Delta_{\mathrm{c}}^{2}}$ 
satisfies $g_{\mathrm{sc}}(r_{0}^{\mathrm{sc}})=0$, 
we have the inequality 
$g_{\mathrm{sc}}(r)
>g_{\mathrm{sc}}(r_{0}^{\mathrm{sc}})=0$ 
provided that $r_{0}^{\mathrm{sc}}<r$.
This fact tells us that 
\begin{align}
& \left(\frac{|\Omega(z,t)|}{
|\Delta_{\mathrm{c}}|}\right)^{2}
>
m+n 
+2\sqrt{
mn+\frac{K^{2}}{4\Delta_{\mathrm{c}}^{2}}\,} \notag \\ 
\Longrightarrow\,\,\, 
& 
g_{\mathrm{sc}}\left(
\left(\frac{|\Omega(z,t)|}{
|\Delta_{\mathrm{c}}|}\right)^{2}\right)
> 0.
\label{eq:081014-5'}
\end{align}
Therefore, we can conclude the inequality 
(\ref{eq:after-DC}) from implications 
(\ref{eq:081014-4'}) and (\ref{eq:081014-5'}). 

In the case where $K<0$, we have 
\begin{align*}
& E_{m}^{-}(z,t;1)-E_{n}^{-}(z,t;1)  \notag \\ 
=& (n-m)\left[
\Delta_{\mathrm{c}} 
+ 
{\displaystyle \frac{\Omega(z,t)^{2}}{
\Upsilon_{n+1}(z,t;;1)
+\Upsilon_{m+1}(z,t;;1)}}
\right].
\end{align*}
Hence it follows from this that 
we obtain the crossing (\ref{eq:before-DC}) 
and (\ref{eq:after-DC}) 
by using $m+1$ and $n+1$ instead of $m$ and 
$m$ respectively in the above argument for 
the case where $K\ge 0$.

\subsection{In the case $d \ge 2$}
\label{subsec:d>=2}

We assume $d \ge 2$ in this subsection. 
Let $m$ and $n$ be non-negative integers 
satisfying $m < d \le n$. 
We set a positive 
number $C_{mn}^{0}(d)$ 
by 
$$
C_{mn}^{0}(d) 
:=
\begin{cases} 
{\displaystyle \frac{(n-d)!}{n!}}
(n-m)|\Delta_{\mathrm{c}}|
\Bigl\{
|\Delta_{\mathrm{c}}|(n-m)
+K_{d}
\Bigr\} 
& \quad \\ 
\qquad\qquad\qquad\qquad\text{if $\varepsilon_{1}-\varepsilon_{0}
\ge \Delta-d\Delta_{\mathrm{c}}$}, 
& \quad \\ 
{\displaystyle \frac{n!}{(n+d)!}}
(n+d-m)|\Delta_{\mathrm{c}}|\times 
& \quad \\ 
\qquad\qquad\qquad 
\times 
\Bigl\{
|\Delta_{\mathrm{c}}|(n+d-m)
+K_{d}
\Bigr\} 
& \quad \\ 
\qquad\qquad\qquad\qquad\text{if $\varepsilon_{1}-\varepsilon_{0}
< \Delta-d\Delta_{\mathrm{c}}$},
& \quad 
\end{cases}
$$ 
where $K_{d}:=
\varepsilon_{1}-\varepsilon_{0}
+d\Delta_{\mathrm{c}}-\Delta$. 
Then, we define three domains 
$\mathcal{D}^{\mathrm{wc}}_{mn}(d)$, 
$\mathcal{D}^{\mathrm{sc}}_{mn}(d)$, and 
$\mathcal{D}^{\mathrm{cr}}_{mn}(d)$ 
of the space-time in the following: 
the domain $\mathcal{D}_{mn}^{\mathrm{wc}}(d)$ 
for the weak coupling regime is given by
$\mathcal{D}^{\mathrm{wc}}_{mn}(d):= 
\left\{
(z,t) \, |\, 
|\Omega(z,t)|^{2} < C_{mn}^{0}
\right\}$, 
and the domain $\mathcal{D}_{mn}^{\mathrm{sc}}(d)$ 
for the strong coupling regime by 
$\mathcal{D}^{\mathrm{sc}}_{mn}(d):= 
\left\{
(z,t) \, |\, 
|\Omega(z,t)|^{2} > C_{mn}^{0}(d)
\right\}$. 
The domain $\mathcal{D}_{mn}^{\mathrm{cr}}(d)$ 
for the critical regime is defined by 
$\mathcal{D}^{\mathrm{cr}}_{mn}(d):= 
\left\{
(z,t) \, |\, 
|\Omega(z,t)|^{2} = C_{mn}^{0}(d)
\right\}$. 

In the case $d\ge 2$ 
we can show the following theorem: 
\textit{the domain $\mathbb{D}^{\mathrm{wc}}_{mn}(d)$ 
is equal to the domain 
$\mathcal{D}^{\mathrm{wc}}_{mn}(d)$, 
and 
the domain $\mathbb{D}^{\mathrm{sc}}_{mn}(d)$ 
to the domain $\mathcal{D}^{\mathrm{sc}}_{mn}(d)$, 
i.e.,  
$\mathbb{D}^{\mathrm{wc}}_{mn}(d)
=\mathcal{D}^{\mathrm{wc}}_{mn}(d)$ 
and $\mathbb{D}^{\mathrm{sc}}_{mn}(d)
=\mathcal{D}^{\mathrm{sc}}_{mn}(d)$, 
for every $m$ and $n$ with 
$m < d \le n$. 
Namely, the energy level crossing takes place as}
\begin{align}
& E_{m}^{0}(z,t;d)<E_{n}^{-}(z,t;d)\,\,\, 
\textit{if and only if $(z,t)$ in 
$\mathcal{D}^{\mathrm{wc}}_{mn}(d)$}, 
\label{eq:before-DCd} \\ 
& E_{m}^{0}(z,t;d)=E_{n}^{-}(z,t;d)\,\,\, 
\textit{if and only if $(z,t)$ in 
$\mathcal{D}^{\mathrm{cr}}_{mn}(d)$}, 
\label{eq:0-DCd} \\ 
& E_{m}^{0}(z,t;d)>E_{n}^{-}(z,t;d)\,\,\, 
\textit{if and only if $(z,t)$ in 
$\mathcal{D}^{\mathrm{sc}}_{mn}(d)$}. 
\label{eq:after-DCd} 
\end{align}
Here we note 
\begin{align*}
 C_{0n}^{0}(d)  
=&  
\Delta_{\mathrm{c}}^{2}
\frac{n}{(n-1)\cdots(n-d+1)} \\ 
& + 
|\Delta_{\mathrm{c}}|
\frac{K_{d}}{(n-1)\cdots(n-d+1)} \\ 
<& 
\Delta_{\mathrm{c}}^{2}n
+ 
|\Delta_{\mathrm{c}}|K
\le C_{0n}^{0} 
\end{align*}
because $d\Delta_{\mathrm{c}} < \Delta_{\mathrm{c}} < 0$. 
Thus, it follows from this inequality 
that $\mathcal{D}^{\mathrm{sc}}_{0n}(1)
\subset \mathcal{D}^{\mathrm{sc}}_{0n}(d)$. 
It means that \textit{non-linear coupling} 
(\textit{i.e., $d \ge 2$}) \textit{causes 
the Dicke-type energy level 
crossing easier than linear coupling} 
(\textit{i.e., $d=1$}) does. 

Before proving our desired statements, 
we set $K_{d}$ as $K_{d}:= 
\varepsilon_{1}-\varepsilon_{0}
+d\Delta_{\mathrm{c}}-\Delta$ 
for the simplicity in calculation. 
It is easy to check that 
\begin{align*}
&E_{m}^{0}(z,t;d)-E_{n}^{-}(z,t;d) \\ 
=&
\begin{cases}
-\Delta_{\mathrm{c}}(m-n)
- {\displaystyle \frac{K_{d}}{2}}
+\Upsilon_{n}(z,t;d) 
& \text{if $K_{d}\ge 0$},  \\ 
-\Delta_{\mathrm{c}}(m-n-d)
-{\displaystyle \frac{K_{d}}{2}}
+\Upsilon_{n+d}(z,t;d)
& \text{if $K_{d}<0$}, 
\end{cases}
\end{align*}
which implies the following equivalence: 
\begin{align}
&E_{m}^{0}(z,t;d)-E_{n}^{-}(z,t;d)\,\,\sharp\,\, 0  \notag \\ 
\Longleftrightarrow & 
\begin{cases}
\Upsilon_{n}(z,t;d)
\,\,\sharp\,\, 
-\Delta_{\mathrm{c}}(n-m)
+{\displaystyle \frac{K_{d}}{2}}
& \quad \\ 
\qquad\qquad\qquad\qquad\qquad
\text{if $K_{d}\ge 0$},  \\ 
\Upsilon_{n+d}(z,t;d)
\,\,\sharp\,\, 
-\Delta_{\mathrm{c}}(n+d-m)
+{\displaystyle \frac{K_{d}}{2}}
& \quad \\ 
\qquad\qquad\qquad\qquad\qquad\qquad
\text{if $K_{d}<0$}.
\end{cases}
\label{eq:081017-0}
\end{align}

We see the energy level crossing 
in the case where $K_{d}\ge 0$ first. 
Thus, suppose $K_{d}\ge 0$. 
Since $|\Delta_{\mathrm{c}}|=-\Delta_{\mathrm{c}}$, 
multiplying both sides of the expression 
$|\Omega(z,t)|^{2}\,\,\sharp\,\, C_{mn}^{0}(d)$ 
by $n!/(n-d)!$, 
we know that the expression is equivalent to 
the expression 
$|\Omega(z,t)|^{2}n!/(n-d)!\,\,\sharp\,\, 
\Delta_{\mathrm{c}}^{2}(n-m)^{2}
-\Delta_{\mathrm{c}}(n-m)K_{d}$. 
Adding $K_{d}^{2}/4$ to both sides of the 
multiplied expression and factorizing 
the left hand side with $1/4$, 
we know the expression 
is equivalent to the expression 
$(K_{d}^{2}
+4|\Omega(z,t)|^{2}n!/(n-d)!)/4\,\,\sharp\,\, 
\Delta_{\mathrm{c}}^{2}(n-m)^{2}
-\Delta_{\mathrm{c}}(n-m)K_{d}+K_{d}^{2}/4$. 
Taking the square root of 
both sided of this expression, 
we obtain the equivalence: 
\begin{align}
& |\Omega(z,t)|^{2}\,\,\sharp\,\, 
C_{mn}^{0}(d) \notag  \\ 
\Longleftrightarrow  & 
\Upsilon_{n}(z,t;1)\,\,\sharp\,\,  
-\Delta_{\mathrm{c}}(n-m)
+\frac{K_{d}}{2}
\label{eq:081017-1}
\end{align}
since $\Delta_{\mathrm{c}}<0$, 
$m<n$, and $0\le K_{d}$.  
Finally, by the equivalences 
(\ref{eq:081017-0}) and 
(\ref{eq:081017-1}), 
we reach the equivalence: 
\begin{equation}
E_{m}^{0}(z,t;d)-E_{n}^{-}(z,t;d)\,\,\sharp\,\, 0 
\Longleftrightarrow 
|\Omega(z,t)|^{2}\,\,\sharp\,\, 
C_{mn}^{0}(d),
\label{eq:081017-2}
\end{equation}
which says that the desired energy level 
crossing takes place when $K_{d}\ge 0$. 

We see the energy level crossing 
in the case where $K_{d}< 0$ next.
Let us suppose $K_{d}<0$ now. 
Note the inequalities 
$-\Delta_{\mathrm{c}}(n+d-m)+K_{d}/2
\ge (\varepsilon_{1}-\varepsilon_{0}
-d\Delta_{\mathrm{c}}-\Delta)/2 > 
(\varepsilon_{1}-\varepsilon_{0}
-\Delta_{\mathrm{c}}-\Delta)/2 > 
(\varepsilon_{1}-\varepsilon_{0}
-\Delta)/2>0$ because of the assumption 
(\ref{eq:assumption}) and the condition 
$-\Delta_{\mathrm{c}}>0$.  
Thus, in the same way we had the equivalence 
(\ref{eq:081017-1}), we obtain 
the equivalence, 
\begin{align}
& |\Omega(z,t)|^{2}\,\,\sharp\,\, 
C_{mn}^{0}(d) \notag \\ 
\Longleftrightarrow & 
\Upsilon_{n+d}(z,t;d)\,\,\sharp\,\, 
-\Delta_{\mathrm{c}}(n+d-m)
+\frac{K_{d}}{2}.
\label{eq:081017-1'}
\end{align}
Then, the equivalences 
(\ref{eq:081017-0}) and 
(\ref{eq:081017-1'}) bring 
the equivalence (\ref{eq:081017-2}), 
which secures our statement 
in the case $K_{d}<0$.

\textit{In the case where $d\ge 2$, 
there is not always 
a ground state of $H_{0}(z,t;d)$. 
Namely, there is a case that the minimum 
energy of $H_{0}(z,t;d)$ 
does not exist.} 
To see this fact, we assume $\varepsilon_{1}-
\varepsilon_{0}\approx \Delta - d\Delta_{\mathrm{c}}$ 
for simplicity. 
Then, since we have 
\begin{align*}
& E_{n}^{-}(z,t;d) - E_{n+1}^{-}(z,t;d) \\ 
&= 
\Delta_{\mathrm{c}} 
+ |\Omega(z,t)|
\sqrt{n(n-1)\cdots (n-d+2)}\times \\ 
&\qquad\qquad\qquad\qquad\qquad \times 
\left\{
\sqrt{n+1}-\sqrt{n-d+1} 
\right\} \\ 
&> 
\Delta_{\mathrm{c}} 
+ |\Omega(z,t)|
\sqrt{n(n-1)\cdots (n-d+2)}\times \\ 
&\qquad\qquad\qquad\qquad\qquad \times
\left\{
\sqrt{n+1}-\sqrt{n-1} 
\right\} \\ 
&= 
\Delta_{\mathrm{c}} 
+ 2|\Omega(z,t)|
\frac{\sqrt{(n-1)\cdots (n-d+2)}}{
\sqrt{1+1/n}+\sqrt{1-1/n}} \\ 
&> 
\Delta_{\mathrm{c}} 
+ |\Omega(z,t)|
\sqrt{(d-1)!},  
\end{align*}
where we used the inequalities, 
$n> d \ge 2$ and 
$\sqrt{1+1/n}+\sqrt{1-1/n} \ge 2$.    
The last inequality says that 
\textit{the Hamiltonian $H_{0}(z,t;d)$ 
does not have a ground state for 
$(z,t)$ satisfying $|\Omega(z,t)| > 
|\Delta_{\mathrm{c}}|/\sqrt{(d-1)!}$
because} 
$$
\cdots < E_{n+1}^{-}(z,t;d) < E_{n}^{-}(z,t;d) 
< \cdots < E_{d+1}^{-}(z,t;d)
$$   
\textit{in the case where $d\ge 2$}.

\section{Superradiant Ground State 
Energy in The Case $\alpha\equiv 0$.}
\label{sec:sgse}

As we knew at the end of the previous section, 
once the Dicke-type crossing takes place 
in the case $d \ge 2$, 
there is \textit{every} possibility that 
$H_{0}(z,t;d)$ is not bounded from below. 
Thus, to make sure of the existence of 
the superradiant ground state energy, 
we consider only the case $d=1$ in this section.

Set a positive number $\Theta_{n}$ 
for $n = 1, 2, \cdots$ as 
$$
\Theta_{n} = 
\begin{cases}
2n\Delta_{\mathrm{c}}^{2} 
+ |\Delta_{\mathrm{c}}|
\sqrt{4n^{2}\Delta_{\mathrm{c}}^{2} 
+ K^{2}}  
& \quad \\ 
\qquad\qquad\qquad\qquad\qquad
\text{if $\varepsilon_{1}-\varepsilon_{0}
\ge \Delta-\Delta_{\mathrm{c}}$,} 
& \quad \\ 
2(n+1)\Delta_{\mathrm{c}}^{2}
+ |\Delta_{\mathrm{c}}|
\sqrt{4(n+1)^{2}\Delta_{\mathrm{c}}^{2} 
+ K^{2}} 
& \qquad \\ 
\qquad\qquad\qquad\qquad\qquad
\text{if $\varepsilon_{1}-\varepsilon_{0}
< \Delta-\Delta_{\mathrm{c}}$,} 
& \quad 
\end{cases}
$$
where 
$K=\varepsilon_{1}-\varepsilon_{0}
+\Delta_{\mathrm{c}}-\Delta$. 
Using this number, we define 
a domain $\mathcal{G}_{n}(1)$ of the time-space 
by 
$$
\mathcal{G}_{n}(1) = 
\left\{
(z,t)\, \bigg|\, 
\sqrt{\Theta_{n}} < |\Omega(z,t)| 
< \sqrt{\Theta_{n+1}}
\right\}
$$
for each $n = 1, 2, \cdots$. 
We can prove the following 
theorem: 
\textit{(i) when a space-time 
point $(z,t)$ is in $\mathcal{G}_{n}(1)$, 
the point $(z,t)$ is always 
in $\mathcal{D}_{0n}^{\mathrm{sc}}(1)$. 
Namely, the Dicke-type energy level 
crossing takes place 
for that point $(z,t)$ in $\mathcal{G}_{n}(1)$; 
(ii) the superradiant ground state energy 
$\inf\mathrm{Spec}
(H_{0}(z,t;1))$ appears as:} 
\begin{equation}
\inf\mathrm{Spec}
(H_{0}(z,t;1)) 
= 
\min\{ E_{n}^{-}(z,t;1) ,  
E_{n+1}^{-}(z,t;1)\} 
\label{eq:superradiant-gse}
\end{equation}
for $(z,t)$ in  $\mathcal{G}_{n}(1)$. 

To prove our theorem we set $K$ 
as $K:=\varepsilon_{1}-\varepsilon_{0}
+\Delta_{\mathrm{c}}-\Delta$ again.  
Part (i) follows from the following 
easy inequalities: 
$C_{0n}^{0} 
< 2n\Delta_{\mathrm{c}}^{2} 
+|\Delta_{\mathrm{c}}|K \le \Theta_{n}$ 
if $K\ge 0$; 
$C_{0n}^{0} 
< 2(n+1)\Delta_{\mathrm{c}}^{2} 
+|\Delta_{\mathrm{c}}|K \le \Theta_{n}$ 
if $K< 0$.
For a non-negative number $r$, 
define a function 
$g(r)$ by $g(r):=|\Delta_{\mathrm{c}}|r+L
-(1/2)\sqrt{K^{2}+4|\Omega(z,t)|^{2}r\,}$, 
where $L=(\varepsilon_{0}+\varepsilon_{1}
+\Delta_{\mathrm{c}}-\Delta)$. 
Then, for every positive number $r$ 
the expression $g'(r)\,\,\sharp\,\, 0$ 
is equivalent to the expression 
$\Delta_{\mathrm{c}}^{2}(K^{2}
+4|\Omega(z,t)|^{2}r)
\,\,\sharp\,\, |\Omega(z,t)|^{4}$. 
Hence it follows from this that 
\begin{equation}
g'(r)\,\,\sharp\,\, 0 
\Longleftrightarrow 
r\,\,\sharp\,\, 
r_{0}:= 
\frac{|\Omega(z,t)|^{2}}{4\Delta_{\mathrm{c}}^{2}} 
- 
\frac{K^{2}}{4|\Omega(z,t)|^{2}},  
\label{eq:081020-1}
\end{equation}
provided that $r_{0}>0$. 
It is evident that the number $r_{0}$ is 
positive if and only if 
the inequality $|\Omega(z,t)|^{2} 
> |\Delta_{\mathrm{c}}K|$ holds. 
By the equivalence (\ref{eq:081020-1}) 
together with this fact, we have the implication: 
\begin{equation}
|\Omega(z,t)|^{2} > |\Delta_{\mathrm{c}}K| 
\Longrightarrow  
\inf_{r\ge 0}g(r)=g(r_{0}).   
\label{eq:081020-3}
\end{equation}
Simple calculation leads to the equivalence: 
\begin{align}
& n 
\le 
\frac{|\Omega(z,t)|^{2}}{4\Delta_{\mathrm{c}}^{2}} 
- \frac{K^{2}}{4|\Omega(z,t)|^{2}} < n+1 
\notag \\ 
\Longleftrightarrow & 
\begin{cases}
0 \le |\Omega(z,t)|^{4} 
- 4n\Delta_{\mathrm{c}}^{2}|\Omega(z,t)|^{2} 
- \Delta_{\mathrm{c}}^{2}K^{2}, & \quad \\ 
\qquad \\ 
|\Omega(z,t)|^{4} 
- 4(n+1)\Delta_{\mathrm{c}}^{2}|\Omega(z,t)|^{2} 
- \Delta_{\mathrm{c}}^{2}K^{2} <0. 
& \quad 
\end{cases}
\label{eq:081020-5}
\end{align}
We define two functions $g_{j}(r)$, $j=1, 2$, 
for positive number $r$ by 
$g_{1}(r):=r^{2}-4n\Delta_{\mathrm{c}}^{2}r 
-\Delta_{\mathrm{c}}^{2}K^{2}$ 
and 
$g_{2}(r):=r^{2}-4(n+1)\Delta_{\mathrm{c}}^{2}r 
-\Delta_{\mathrm{c}}^{2}K^{2}$. 
Then, setting $r_{j0}$ as 
$r_{10}:=2n\Delta_{\mathrm{c}}^{2}
+|\Delta_{\mathrm{c}}|
\sqrt{K^{2}+4n^{2}\Delta_{\mathrm{c}}^{2}}$ 
and 
$r_{20}:=2(n+1)\Delta_{\mathrm{c}}^{2}
+|\Delta_{\mathrm{c}}|
\sqrt{K^{2}+4(n+1)^{2}\Delta_{\mathrm{c}}^{2}}$ 
respectively, 
we have 
$g_{1}(r)>g_{1}(r_{10})=0$ if $r>r_{10}$ 
and 
$g_{2}(r)<g_{2}(r_{20})=0$ if $0< r<r_{20}$.  
Set a non-negative number $\theta_{n}$ 
for each $n=1, 2, \cdots$ as 
$\theta_{n}:= 
2n\Delta_{\mathrm{c}}^{2}
+|\Delta_{\mathrm{c}}|
\sqrt{K^{2}+4n^{2}\Delta_{\mathrm{c}}^{2}}$. 
Then, we have 
$\theta_{n}\ge |\Delta_{\mathrm{c}}K|$
So, inserting $|\Omega(z,t)|^{2}$ into $r$ 
of the above inequalities, 
we reach the implication: 
\begin{align}
& 
\theta_{n} < |\Omega(z,t)|^{2} < \theta_{n+1}  \notag \\ 
\Longrightarrow &  
\text{
RHS of the equivalence 
(\ref{eq:081020-5}) is satisfied}.   
\label{eq:081020-6}
\end{align}
Combining the results (\ref{eq:081020-3}), 
(\ref{eq:081020-5}), and (\ref{eq:081020-6}), 
we obtain the implication:  
\begin{align}
& 
\theta_{n} < |\Omega(z,t)|^{2} < \theta_{n+1}  
\notag \\ 
\Longrightarrow &
\text{$g(n)$ or $g(n+1)$ 
is located nearest 
$\inf_{r\ge 0}g(r)$}.   
\label{081020-7}
\end{align}
Noting $E_{n}^{-}(z,t;1)=g(n)$ if $K\ge 0$; 
$E_{n}^{-}(z,t;1)=g(n+1)$ if $K<0$, and 
$\Theta_{n}=\theta_{n}$ if $K\ge 0$; 
$\Theta_{n}=\theta_{n+1}$ if $K<0$, 
we obtain part (ii).

\section{Stability of Dicke-Type Crossings 
in The Case $\alpha\equiv\!\!\!\!\!\!{/}\,\, 0$.}
\label{sec:stability}

In this section we consider the case where 
$d=1$ too. 
we show the stability of 
the Dicke-type energy level crossings 
under the effect of the generalized 
energy operator $\alpha(t)W(z,t)$ 
of the pump field for sufficiently 
small strength $|\alpha(t)|$. 
Our $W(z,t)$ includes $i\alpha(a-a^{\dagger})$, 
of course.
To show this fact we employ the method for proving 
Theorem 4.3(v) of Ref.\onlinecite{hir-iumj}.
In the same way we did in Sec.\ref{sec:dicke-type-crossing}, 
through the well-known identification 
as in Ref.\onlinecite{H-R}, 
$H(\Omega,\alpha;1)$ reads 
$H_{\alpha}(z,t;1):=H_{0}(z,t;1)+\alpha(z,t)W(z,t)$. 
We recall $\alpha(z,0)=0$ for all the position $z$. 

For the Hilbert space representing the state space 
in which $H_{\alpha}(z,t;d)$ acts, 
we denote the inner product of the Hilbert space 
by $\langle\Psi|\Phi\rangle$. 

For every positive number $\epsilon$, 
we denote $C_{\epsilon}(z,t)$ 
by 
\begin{align*}
& C_{\epsilon}(z,t) \\ 
:=&
(1+\epsilon)
\Biggl\{
1+\left(\frac{\epsilon}{1+\epsilon}\right)^{2}
\left(\varepsilon_{0}+|\varepsilon_{1}-\Delta|\right)
 \notag \\ 
&\qquad\qquad\qquad\qquad\qquad 
+ \left(\frac{1+\epsilon}{\epsilon}\right)^{2}
\frac{|\Omega(z,t)|^{2}}{4\Delta_{\mathrm{c}}^{2}}
\,\Biggr\}^{1/2}.
\end{align*}
Set a positive number $C_{0n}^{0}[\theta]$ for each 
natural number $n$ and a non-negative number $\theta$ as:
\begin{align*}
&C_{0n}^{0}[\theta] \\ 
:=&
\begin{cases}
(\theta-\Delta_{\mathrm{c}})^{2}n
+(\theta-\Delta_{\mathrm{c}})
(\varepsilon_{1}-\varepsilon_{0}+\Delta_{\mathrm{c}}-\Delta) 
& \quad \\ 
\qquad\qquad\qquad\qquad\qquad\qquad
\text{if $\varepsilon_{1}-\varepsilon_{0}\ge 
\Delta-\Delta_{\mathrm{c}}$}, \\
(\theta-\Delta_{\mathrm{c}})^{2}(n+1)
+(\theta-\Delta_{\mathrm{c}})
(\varepsilon_{1}-\varepsilon_{0}+\Delta_{\mathrm{c}}-\Delta) 
& \quad \\ 
\qquad\qquad\qquad\qquad\qquad\qquad
\text{if $\varepsilon_{1}-\varepsilon_{0}< 
\Delta-\Delta_{\mathrm{c}}$}, 
\end{cases}
\end{align*}
and set a positive number $C[n]$ for each 
natural number $n$ as: 
\begin{align*}
& C[n] \\ 
:= &
\begin{cases}
\Delta_{\mathrm{c}}^{2}n 
+{\displaystyle 
\frac{\varepsilon_{0}(\varepsilon_{1}
+\Delta_{\mathrm{c}}-\Delta)}{n}} 
& \quad \\ 
\qquad\qquad 
- \Delta_{\mathrm{c}}(\varepsilon_{0}
+\varepsilon_{1}+\Delta_{\mathrm{c}}-\Delta) 
& \quad \\ 
\qquad\qquad\qquad\qquad\qquad\qquad
\text{if $\varepsilon_{1}-\varepsilon_{0}\ge 
\Delta-\Delta_{\mathrm{c}}$}, \\ 
\Delta_{\mathrm{c}}^{2}(n+1) 
+{\displaystyle 
\frac{\varepsilon_{0}(\varepsilon_{1}
+\Delta_{\mathrm{c}}-\Delta)}{n+1}} 
& \quad \\ 
\qquad\qquad 
- \Delta_{\mathrm{c}}(\varepsilon_{0}
+\varepsilon_{1}+\Delta_{\mathrm{c}}-\Delta) 
& \quad \\ 
\qquad\qquad\qquad\qquad\qquad\qquad
\text{if $\varepsilon_{1}-\varepsilon_{0}< 
\Delta-\Delta_{\mathrm{c}}$}. 
\end{cases}
\end{align*}
Moreover, for every positive function $f(z,t)$, 
and positive numbers $b$ and $\epsilon'$, 
we define a domain $\mathcal{D}(\epsilon,\epsilon';b,f)$ 
of the space-time by 
\begin{align*}
& \mathcal{D}(\epsilon,\epsilon';b,f) \\ 
:=&
\Biggl\{
(z,t)\, \Bigg|\, 
|\alpha(z,t)| <\frac{\epsilon'}{2b}\,\,\,\,\,
\textrm{and} \\ 
&\qquad\qquad 
|\alpha(z,t)|\biggl( 
bC_{\epsilon}(z,t) 
+f(z,t)\biggr) <
\frac{\epsilon'(1+\epsilon)}{2}
\Biggr\}.
\end{align*} 
Also the domain $\mathcal{D}_{0n}^{\mathrm{sc}}(1;\theta)$ 
of the space-time is defined by 
\begin{align*}
& \mathcal{D}_{0n}^{\mathrm{sc}}(1;\theta) \\ 
:= &
\left\{ (z,t)\, |\, 
|\Omega(z,t)|^{2}>C_{0n}^{0}[\theta]\,\,\,\text{and}\,\,\,
|\Omega(z,t)|^{2}>C[n]\right\}. 
\end{align*}

We prove the following theorem concerning 
the stability of the Dicke-type energy level 
crossings:
\textit{Let our $W(z,t)$ satisfy 
the conditions (A1)--(A3):} 
\begin{enumerate}
\item[\textit{(A1)}] 
\textit{For every space-time point 
$(z,t)$, $W(z,t)$ is a symmetric operator 
so that it can act the all states on which 
$H_{0}(z,0;1)$ acts;} 
\item[\textit{(A2)}] 
\textit{there is a positive constant 
$b_{1}$ so that} 
\begin{align}
& \langle W(z,t)\Psi|W(z,t)\Psi\rangle^{1/2} 
\notag \\ 
\le & 
b_{1}\langle H_{0}(z,0;1)\Psi|
H_{0}(z,0;1)\Psi\rangle^{1/2}
+b_{2}(z,t)\langle \Psi|\Psi\rangle^{1/2}
\label{assump:A2}
\end{align}
\textit{for all states $\Psi$ on which 
$H_{0}(z,0;1)$ acts 
and every space-time point $(z,t)$, 
where $b_{2}(z,t)$ is some positive function 
of $(z,t)$;}
\item[\textit{(A3)}] 
\textit{there is a constant $\epsilon$ 
with $0<\epsilon<1$ so that 
$|\alpha(z,t)|<\left\{ 
b_{1}(1+\epsilon)
\right\}^{-1}$ for all the space-time points 
$(z,t)$.} 
\end{enumerate}
\textit{Then, the Hamiltonian $H_{\alpha}(z,t;1)$ 
has an eigenvalue $\mathcal{E}_{n}^{\natural}(z,t;1)$ 
near $E_{n}^{\natural}(z,t;1)$ for each non-negative 
integer $n$, where $\natural=0,\pm$. 
Moreover, there is a constant $\kappa_{0}$ 
with $0<\kappa_{0} <1/4$ so that 
the Dicke-type energy level crossing takes place 
between the eigenvalues 
$\mathcal{E}_{0}^{0}(z,t;1)$ 
and $\mathcal{E}_{n}^{-}(z,t;1)$  
in the process from the space-time point 
$(z,0)$ to the space-time point $(z_{*},t_{*})$, 
provided that the latter point $(z_{*},t_{*})$ 
is in $\mathcal{D}(\epsilon,\kappa_{0};b_{1},b_{2})
\cap \mathcal{D}_{0n}(1;\theta)$ 
for a number $\theta$ with $\theta >0$. 
Therefore, $H_{\alpha}(z_{*},t_{*};1)$ has 
the superradiant ground state energy.}

We prove this theorem below.
Set $H_{\mathrm{I}}$ as 
$H_{\mathrm{I}}:=H_{0}(z,t;1)-H_{0}(z,0;1)$ 
for simplicity. 
The symbol $H_{0}$ stands for 
$H_{0}(z,0;1)$ from now on.
Here we recall $\Omega(z,0)=0$ for every position $z$. 

In the same way that we proved Lemma 4.2 of 
Ref.\onlinecite{hir-iumj}, 
we can estimate $\langle W(z,t)\Psi|
W(z,t)\Psi\rangle$ from above 
by using $\langle H_{0}(z,t;d)\Psi|H_{0}(z,t;d)\Psi\rangle$ 
and $\langle \Psi|\Psi\rangle$. 
Using the canonical commutation relation 
$[a,a^{\dagger}]=1$ leads to the equation: 
$$
\langle H_{\mathrm{I}}\Psi|
H_{\mathrm{I}}\Psi\rangle =
|\Omega(z,t)|^{2}\left(
\langle\Psi|a^{\dagger}a|\Psi\rangle+
\langle\sigma_{11}\Psi|\sigma_{11}\Psi\rangle\right),
$$ 
which easily implies the inequality 
$$
\langle H_{\mathrm{I}}\Psi|H_{\mathrm{I}}\Psi\rangle 
\le 
|\Omega(z,t)|^{2}\left(
\langle\Psi|a^{\dagger}a|\Psi\rangle 
+\langle\Psi|\Psi\rangle\right).
$$ 
By using the Schwarz inequality and the inequality 
$XY\le (\eta X+Y/4\eta)^{2}$ for every $X, Y \ge 0$ 
and arbitrary $\eta > 0$, 
we obtain the inequality:
$$
\langle\Psi|a^{\dagger}a|\Psi\rangle 
\le 
\left(\eta\langle 
a^{\dagger}a\Psi|a^{\dagger}a\Psi\rangle^{1/2} 
+\langle\Psi|\Psi\rangle^{1/2}/(4\eta)\right)
$$ 
for arbitrary $\eta > 0$. 
Simple calculation yields 
\begin{align*}
&\langle H_{0}\Psi|H_{0}\Psi\rangle \\ 
=& 
\Delta_{\mathrm{c}}^{2}
\langle a^{\dagger}a\Psi|a^{\dagger}a\Psi\rangle 
+(\varepsilon_{1}-\Delta)
\langle\sigma_{11}\Psi|\sigma_{11}\Psi\rangle \\ 
& +
\varepsilon_{0}
\langle\sigma_{00}\Psi|\sigma_{00}\Psi\rangle,
\end{align*}
so that 
\begin{align*}
&
\langle a^{\dagger}a\Psi|a^{\dagger}a\Psi\rangle \\ 
=& 
(1/\Delta_{\mathrm{c}}^{2})
\Biggl\{
\langle H_{0}\Psi|H_{0}\Psi\rangle
-(\varepsilon_{1}-\Delta)
\langle\sigma_{11}\Psi|\sigma_{11}\Psi\rangle \\ 
&\qquad\qquad 
-\varepsilon_{0}
\langle\sigma_{00}\Psi|\sigma_{00}\Psi\rangle
\Biggr\}.
\end{align*}  
Combining these inequalities and setting $\eta$ 
as $\eta:=|\Delta_{\mathrm{c}}|\delta/(\sqrt{2}
|\Omega(z,t)|)$ for arbitrary $\delta > 0$, 
we reach the inequality:  
\begin{align}
&\langle H_{\mathrm{I}}\Psi|
H_{\mathrm{I}}\Psi\rangle^{1/2} \notag \\ 
\le& 
|\Omega(z,t)|
\left\{
2\eta^{2}\langle a^{\dagger}a\Psi |
a^{\dagger}a\Psi\rangle 
+\left(\frac{1}{8\eta^{2}}+1\right)
\langle\Psi |\Psi\rangle
\right\}^{1/2} \notag \\ 
\le& 
\left\{\delta^{2}
\langle H_{0}\Psi|H_{0}\Psi\rangle 
+ 
\Theta(\delta;\Omega)^{2}
\langle\Psi|\Psi\rangle\right\}^{1/2} \notag \\ 
\le& 
\delta
\langle H_{0}\Psi|H_{0}\Psi\rangle^{1/2} 
+\Theta(\delta;\Omega)\langle\Psi|\Psi\rangle^{1/2},
\label{eq:Ineq5}
\end{align}
where 
$$
\Theta(\delta;\Omega)
=
\sqrt{
1+\delta^{2}(\varepsilon_{0}
+|\varepsilon_{1}-\Delta|)
+\frac{|\Omega(z,t)|^{2}}{
4\Delta_{\mathrm{c}}^{2}\delta^{2}}
\,\,}.
$$

Applying the triangle inequality
$\langle (A+B)\Psi|(A+B)\Psi\rangle^{1/2}\le 
\langle A\Psi|A\Psi\rangle^{1/2}
+\langle B\Psi|B\Psi\rangle^{1/2}$ 
to the term $\langle (H_{0}(z,t;1)-H_{\mathrm{I}})\Psi|
(H_{0}(z,t;1)-H_{\mathrm{I}})\Psi\rangle^{1/2}$, 
we have the inequality: 
\begin{align}
\langle H_{0}\Psi|
H_{0}\Psi\rangle^{1/2} 
\le& 
\langle H_{0}(z,t;1)\Psi|H_{0}(z,t;1)\Psi\rangle^{1/2} 
\notag \\ 
&+ \langle H_{\mathrm{I}}
\Psi|H_{\mathrm{I}}\Psi\rangle^{1/2}. 
\label{eq:Ineq6}
\end{align}
Inequalities (\ref{eq:Ineq5}) and (\ref{eq:Ineq6}) 
tell us that the term $\langle H_{0}\Psi|
H_{0}\Psi\rangle^{1/2}$ is bounded from above as:  
\begin{align*}
\langle H_{0}\Psi|
H_{0}\Psi\rangle^{1/2} 
\le&  
\langle H_{0}(z,t;1)\Psi|H_{0}(z,t;1)\Psi\rangle^{1/2} \\ 
&
+\delta\langle H_{0}\Psi|
H_{0}\Psi\rangle^{1/2} 
+\Theta(\delta;\Omega)
\langle\Psi|\Psi\rangle^{1/2},
\end{align*}
which implies the inequality,  
\begin{align}
(1-\delta)\langle H_{0}\Psi|
H_{0}\Psi\rangle^{1/2} 
\le& 
\langle H_{0}(z,t;1)\Psi|H_{0}(z,t;1)\Psi\rangle^{1/2} 
\notag \\ 
& +\Theta(\delta;\Omega)
\langle\Psi|\Psi\rangle^{1/2}.  
\label{eq:Ineq7}
\end{align}

Set $\delta$ as $0<\delta:=\epsilon(1+\epsilon)^{-1}<1$ 
for arbitrary number $\epsilon$ with $0<\epsilon<1$ now. 
Then, combining the inequalities (\ref{assump:A2}) 
and (\ref{eq:Ineq7}) 
we can conclude that 
\begin{align}
&
\langle W(z,t)\Psi|W(z,t)\Psi\rangle^{1/2}   
\notag \\ 
\le& 
b_{1}(1+\epsilon)
\langle H_{0}(z,t;1)\Psi|H_{0}(z,t;1)\Psi\rangle^{1/2}
\notag \\ 
&
+ (b_{1}C_{\epsilon}(z,t)+b_{2}(z,t))
\langle \Psi|\Psi\rangle^{1/2}. 
\label{eq:Ineq8}
\end{align}
Thus, applying Lemma 4.1 of Ref.\onlinecite{hir-iumj} 
and Theorem 6.29 in III \S{6} of Ref.\onlinecite{kato} 
to our Hamiltonian $H_{\alpha}(z,t;1) =H_{0}+H_{\mathrm{I}}$, 
we know that the Hamiltonian $H_{\alpha}(z,t;1)$ has 
an eigenvalue $\mathcal{E}_{n}^{\natural}(z,t;1)$ 
for each $n=0, 1, 2, \cdots$.

Define an operator $T(\kappa)$ as the closure 
of $H_{0}(z,t;1) + \kappa W(z,t)$ 
for every complex number $\kappa$. 
By applying Theorem 2.6 and Remark 2.7 
in VII\S 2 of Ref.\onlinecite{kato} 
to the inequality (\ref{eq:Ineq8}), 
the operator $T(\kappa)$ is an analytic family of 
type (A) for every $\kappa \in  \mathbb{C}$ with 
$|\kappa| < \left\{b_{1}(1+\epsilon)\right\}^{-1}$.
Thus, taking $\alpha(z,t)$ as this $\kappa$, 
i.e., $H_{\alpha}(z,t;1)=T(\alpha(z,t))$, 
Theorem 3.9 in VII \S{3} of Ref.\onlinecite{kato} 
says that $\mathcal{E}_{n}^{\natural}(z,t;1)$ 
is a continuous function of $(z,t)$ by the 
assumption (A3), and it sits near 
$E_{n}^{\natural}(z,t;1)$. 
So, as shown in Sec.\ref{sec:dicke-type-crossing}, 
candidates which makes the Dicke-type 
energy level crossing are $E^{-}_{n}(z,t;1)$'s. 
Namely, for $H_{\alpha}(z,t;1)$, 
candidates which makes the Dicke-type 
energy level crossing are 
$\mathcal{E}^{-}_{n}(z,t;1)$'s. 

Let $E_{n}$ and $\mathcal{E}_{n}(\kappa)$ be 
respectively eigenvalues 
of $H_{0}(z,t;1)$ and $T(\kappa)$ satisfying 
$\mathcal{E}_{n}(0) = E_{n}$. 
Following the regular perturbation theory 
(see \S XII of Ref.\onlinecite{rs4}), 
the eigenvalue $\mathcal{E}_{n}(\kappa)$ 
has the expression, 
\begin{align}
\mathcal{E}_{n}(\kappa) 
=& 
\frac{\langle\Phi_{n}\, |\, 
T(\kappa)P_{n}(\kappa)\Phi_{n}
\rangle}{
\langle\Phi_{n}\, |\, 
P_{n}(\kappa)\Phi_{n}
\rangle} \notag \\ 
=& 
E_{n} + \kappa 
\frac{\langle\Phi_{n}\, |\, 
W(z,t)P_{n}(\kappa)\Phi_{n}
\rangle}{
\langle\Phi_{n}\, |\, 
P_{n}(\kappa)\Phi_{n}
\rangle}  
\label{eq:perturbation1}
\end{align}
where $P_{n}(\kappa)$ is the orthogonal projection 
given by 
$$
P_{n}(\kappa) = -\, 
\frac{1}{2\pi i}
\oint_{|E-E_{n}|=\widetilde{\epsilon}}
\left( T(\kappa) - E\right)^{-1}dE
$$
with a sufficiently small $\widetilde{\epsilon} > 0$, 
and $\Phi_{n}$ is a normalized eigenvector 
of $H_{0}(z,t;1)$ satisfying 
$H_{0}(z,t;1)\Phi_{n}= E_{n}\Phi_{n}$. 

Since $W(z,t)$ is symmetric, 
use of the Schwarz inequality makes 
the inequality:
\begin{equation}
\Biggl|
\frac{\langle\Phi_{n}\, |\, 
W(z,t)P_{n}(\kappa)\Phi_{n}
\rangle}{
\langle\Phi_{n}\, |\, 
P_{n}(\kappa)\Phi_{n}
\rangle}
\Biggr| 
\le 
\frac{\langle W(z,t)\Phi_{n}|
W(z,t)\Phi_{n}\rangle^{1/2}}{
\langle P_{n}(\kappa)\Phi_{n}|
P_{n}(\kappa)\Phi_{n}
\rangle\,\,}.  
\label{eq:perturbation2}
\end{equation}
To get the right hand side of the above inequality, 
we used the equations 
$P_{n}(\kappa)^{2}=P_{n}(\kappa)
=P_{n}(\kappa)^{*}$ in the denominator, 
and the inequality $\langle P_{n}(\kappa)\Phi_{n}|
P_{n}(\kappa)\Phi_{n}\rangle \le 1$ 
in the numerator. 
By the inequality (\ref{eq:Ineq8}) we can estimate 
$\langle W(z,t)\Phi_{n}|
W(z,t)\Phi_{n}\rangle^{1/2}$ as 
$\langle W(z,t)\Phi_{n}|
W(z,t)\Phi_{n}\rangle^{1/2} 
\le 
b_{1}(1+\epsilon)E_{n} 
+ 
b_{1}C_{\epsilon}(z,t)+b_{2}(z,t)$. 
Here we note that $0\le 
\langle P_{n}(\kappa)\Phi_{n}|
P_{n}(\kappa)\Phi_{n}\rangle 
\longrightarrow 1$ 
as $|\kappa|\to 0$, so that we have 
$1/2<\langle P_{n}(\kappa)\Phi_{n}|
P_{n}(\kappa)\Phi_{n}\rangle$ for sufficiently 
small $|\kappa|$. 
Thus, combining these with the inequality 
(\ref{eq:perturbation2}), 
there is a positive constant $\kappa_{0}$ 
so that 
\begin{equation} 
|\mathcal{E}_{n}(\kappa) - E_{n}| 
\le 2|\kappa| 
\left\{
b_{1}(1+\epsilon)E_{n} 
+ 
b_{1}C_{\epsilon}(z,t)+b_{2}(z,t)
\right\} 
\label{eq:perturbation4}
\end{equation}
if $|\kappa| \le \kappa_{0}$.

Now we take the coupling strength 
$|\alpha(z,t)|$ for 
space-time point $(z,t) \in 
\mathcal{D}(\epsilon,\kappa_{0};b_{1},b_{2})$ 
as the coupling parameter $\kappa$. 
Let us define a positive number 
$\kappa_{1}$ as $\kappa_{1}
:=\kappa_{0}(1+\epsilon)$. 
Then, it is easy to check that 
$2|\alpha(z,t)|b_{1}(1+\epsilon)<\kappa_{1}$ 
and that $2|\alpha(z,t)|\left\{ 
b_{1}C_{\epsilon}(z,t)
+b_{2}(z,t)\right\} <\kappa_{1}$.  
Combining these inequalities with 
the inequality (\ref{eq:perturbation4}), 
we obtain the inequality: 
\begin{equation}
(1-\kappa_{1})E_{n} - \kappa_{1} 
\le 
\mathcal{E}_{n}(\alpha(z,t)) 
\le 
(1+\kappa_{1})E_{n} + \kappa_{1}  
\label{eq:perturbation6}
\end{equation}
for space-time point $(z,t) \in 
\mathcal{D}(\epsilon,\kappa_{0};b_{1},b_{2})$. 
From now on, we take the $\kappa_{0}$ and 
$\epsilon$ so that $0< \kappa_{1} <1/2$.

Let us set a number $L$ as 
$L:=\varepsilon_{0}+\varepsilon_{1}
+\Delta_{\mathrm{c}}-\Delta$ again. 
It is easy to show that 
$\Omega_{n}(z,t;1)^{2}-\Xi_{n}(1)^{2}
=-\Delta_{\mathrm{c}}^{2}n^{2}
+(|\Omega(z,t)|^{2}+\Delta_{\mathrm{c}}^{2}L)n
-\varepsilon_{0}(\varepsilon_{1}+\Delta_{\mathrm{c}}
-\Delta)$, and thus, we obtain the equivalence: 
\begin{align*}
& 
\Omega_{n}(z,t;1)=|\Omega_{n}(z,t;1)|>
|\Xi_{n}(1)|\ge \Xi_{n}(1) \\ 
\Longleftrightarrow &
|\Omega(z,t)|^{2}>\Delta_{\mathrm{c}}^{2}n
-\Delta_{\mathrm{c}}L 
+\frac{\varepsilon_{0}(\varepsilon_{1}+\Delta_{\mathrm{c}}
-\Delta)}{n}. 
\end{align*}
Hence it follows from this that 
$E_{n}^{-}(z,t;1)$ is negative 
for the point 
$(z,t)\in\mathcal{D}_{0n}^{\mathrm{sc}}(1;\theta)$.

In the same way we did to get 
the equivalence (\ref{eq:081207-1}), 
we obtain the equivalence in the following. 
In the case 
$K\equiv 
\varepsilon_{1}-\varepsilon_{0}
+\Delta_{\mathrm{c}}-\Delta$ is non-negative, 
for every $\theta\ge 0$ and 
each natural number $n$ 
we have 
\begin{align}
&
E_{0}^{0}(z,t;1)\,\,\sharp\,\, 
E_{n}^{-}(z,t;1) + n\theta  \notag \\
\Longleftrightarrow &
|\Omega(z,t)|^{2}
\,\,\sharp\,\,  
(\theta-\Delta_{\mathrm{c}})^{2}n 
+(\theta-\Delta_{\mathrm{c}})K
\equiv C_{0n}^{0}[\theta].      
\label{efq:081026-1-1} 
\end{align}
Let $K$ be negative now. 
We note the inequality 
$(\theta-\Delta_{\mathrm{c}})
\left\{(\theta-\Delta_{\mathrm{c}})
(n+1)+K\right\} 
\ge 
(\theta-\Delta_{\mathrm{c}})
\left\{
-\Delta_{\mathrm{c}}+K
\right\}\ge 0$ 
because of the condition (\ref{eq:assumption}). 
Thus, in the same way we did to get 
the equivalence (\ref{eq:081207-1}), 
for every $\theta\ge 0$ and 
each natural number $n$ 
we have 
\begin{align}
&
E_{0}^{0}(z,t;1)\,\,\sharp\,\, 
E_{n}^{-}(z,t;1) + (n+1)\theta \notag \\ 
\Longleftrightarrow &
|\Omega(z,t)|^{2}
\,\,\sharp\,\, 
(\theta-\Delta_{\mathrm{c}})^{2}(n+1) 
+(\theta-\Delta_{\mathrm{c}})K
\equiv C_{0n}^{0}[\theta].    
\label{efq:081026-1-2} 
\end{align}
Thus, we obtain that 
$E_{0}^{0}(z,t;1) > (1+\theta)E_{n}^{-}(z,t;1) + \theta$ 
since 
$1 < 1 + \theta$ and $E_{n}^{-}(z,t;1) < 0$ 
for every point $(z,t) \in 
\mathcal{D}_{0n}^{\mathrm{sc}}(1;\theta)$.
We take the $\theta$ defined by 
the equation 
$\kappa_{1} = \theta/(2+\theta)$ now. 
Then, the following inequality 
holds: 
$E_{0}^{0}(z,t;1) > 
(1+\kappa_{1})E_{n}^{-}(z,t;1)/(1-\kappa_{1}) 
+ 2\kappa_{1}/(1-\kappa_{1})$,  
which implies 
\begin{equation}
(1-\kappa_{1})E_{0}^{0}(z,t;1) 
- \kappa_{1} 
> (1+\kappa_{1})E_{n}^{-}(z,t;1) 
+ \kappa_{1}
\label{eq:8-5}
\end{equation}
for every $(z,t)\in 
\mathcal{D}_{0n}^{\mathrm{sc}}(1;\theta)$.

Combining the inequalities 
(\ref{eq:perturbation6}) and 
(\ref{eq:8-5}) leads to the inequality: 
\begin{align}
\mathcal{E}_{n}^{-}(z,t;1) \le& 
(1+\kappa_{1})E_{n}^{-}(z,t;1) 
+ \kappa_{1} \notag \\ 
<& (1-\kappa_{1})E_{0}^{0}(z,t;1) 
- \kappa_{1} 
< \mathcal{E}_{0}^{0}(z,t;1) 
\label{eq:8-6}
\end{align}
for every $(z,t) \in 
\mathcal{D}(\epsilon,\kappa_{0};b_{1},b_{2})\cap 
\mathcal{D}_{0n}^{\mathrm{sc}}(1;\theta)$.

Since $\mathcal{E}_{n}^{\natural}(z,t;1)$ 
is a continuous function of $(z,t)$ 
and $\mathcal{E}_{0}^{0}(z,0;1) 
=E_{0}^{0}(z,0;1)<E_{n}^{-}(z,0;1)=
\mathcal{E}_{n}^{-}(z,0;1)$, 
the inequality (\ref{eq:8-6}) means 
that the Dicke-type energy level crossing 
takes place. 

\section{Conclusion}
\label{sec:conclusion}

We have showed that the system of a two-level atom 
coupled with a laser in a cavity 
has the Dicke-type energy level crossing in the process 
that the atom-cavity interaction of the system 
undergoes changes between the weak 
coupling regime and the strong one. 
By using the Dicke-type energy level crossing, 
we have found the following two possibilities in 
(mathematical) theory for the cavity-induced atom cooling. 
We can use a laser only for controlling the strength 
of the atom-cavity interaction without 
throwing another laser to the atom for 
driving it to the excited state, 
and moreover, we can obtain much larger 
energy loss caused by cavity decay, 
if we obtain the cavity that implements 
the domain $\mathcal{D}_{0n}^{\mathrm{sc}}(1)$ 
of the space-time. 
Based on these results, 
we can say that we lay mathematical 
foundations for the concept 
of another superradiant cooling 
in addition to that proposed by Domokos and Ritsch.   
Adding the mathematical foundations to 
the idea of the cavity-induced atom cooling 
by Ritsch \textit{et al.}, 
we can also say 
that the process of our superradiant 
cooling requires only cavity decay and 
control of the position of the atom, 
without atomic absorption and emission of 
photons. 
Therefore, whether the mechanism of our 
superradiant cooling can primarily be 
demonstrated or not depends on 
whether we can make such a fine cavity 
that the spatiotemporal domain 
$\mathcal{D}_{01}^{\mathrm{sc}}(1)$ 
for the strong coupling regime  
can be implemented or not 
\cite{HCLK,RBH,MD,MBBBK,TRK,
B-H1,B-H2,C-F,grangier,NCA,IHB}.

\begin{acknowledgments}
This work is partially supported by Japan 
Society for the Promotion of Science (JSPS), Grand-in-Aid for 
Scientific Research (C) 20540171.  
\end{acknowledgments}

\appendix 

\section{The Eigenvalue Problem for $H_{0}(z,t;d)$}
\label{sec:appendix}

In this appendix, to solve the eigenvalue 
problem: $H_{0}(z,t;1)\Psi(g)=E\Psi(g)$, 
we adopt the way that we did in \S 3 and \S 6 
of Ref.\onlinecite{hir-iumj} 
into our calculations. 
Set $\lambda:=i\Omega(z,t)$ for simplicity.  

Let $n<d$ for a while. 
For a complex constant $g$ set 
$\Psi(g):= 
\begin{pmatrix}
0 \\ 
ga^{\dagger\, n}\psi_{0}
\end{pmatrix}$, 
where $\psi_{0}$ is the vacuum state of 
the photon field of our laser. 
It follows immediately that the condition,
$H(d;\Omega,0)\Psi(g)=E\Psi(g)$, 
is equivalent to the condition, $E=\omega n$. 

Let $n\ge d$ now. 
Set 
$
\Psi(g):= 
\begin{pmatrix}
a^{\dagger\, n-d}\psi_{0} \\ 
ga^{\dagger\, n}\psi_{0}
\end{pmatrix}
$ 
this time.
In the same way as in Ref.\onlinecite{hir-iumj}, 
we conclude that the condition,
$H(d;\Omega,0)\Psi(g)=E\Psi(g)$, 
is equivalent to the conditions,
$$
\begin{cases}
(n-d)\omega + \lambda^{*}
\begin{pmatrix}
n \\ 
d
\end{pmatrix}
d!g=E-\mu, \\ 
\lambda+ng\omega=Eg.
\end{cases}
$$ 
Solving these equations, we obtain
$$
g=\frac{d\omega-\mu\pm
\sqrt{(\mu-d\omega)^{2}+
4
\begin{pmatrix}
n\\
d
\end{pmatrix}
d!|\lambda|^{2}\,}}{
2\lambda^{*}
\begin{pmatrix}
n\\
d
\end{pmatrix}
d!}.
$$
and 
$$
E=n\omega+\frac{\mu-d\omega}{2}
\pm\frac{1}{2}
\sqrt{(\mu-d\omega)^{2}+
4
\begin{pmatrix}
n\\
d
\end{pmatrix}
d!\lambda^{2}}.
$$


\begin{thebibliography}{95}



\bibitem{NHTD}
W.~Neuhauser, M.~Hohenstatt, 
P.~E.~Toschek, and H.~Dehmelt,  
Phys. Rev. A \textbf{22}, 1137 (1980).

\bibitem{CHBCA}
S.~Chu, L.~W.~Hollberg, J.~E.~Bjorkholm, 
A.~Cable, and A.~Ashkin,  
Phys. Rev. Lett. \textbf{55}, 48 (1985).


\bibitem{C-TP}
C.~Cohen-Tannoudji and W.~D.~Phillips,  
Physics Today \textbf{43}, 33 (1990).


\bibitem{CW}
E.~A.~Cornell and C.~E.~Wieman, 
Rev. Mod. Phys. \textbf{74}, 875 (2002).


\bibitem{ketterle}
W.~Ketterle, 
Rev. Mod. Phys. \textbf{74}, 1131 (2002).

\bibitem{HCLK}
C.~J.~Hood, M.~S.~Chapman, 
T.~W.~Lynn, and H.~J.~Kimble, 
Phys. Rev. Lett. \textbf{80}, 4157 (1998).

\bibitem{RBH}
J.~M.~Raimond, M.~Brune, and S.~Haroche, 
Rev. Mod. Phys. \textbf{75}, 565 (2001).

\bibitem{MD}
H.~Mabuchi and A.~C.~Doherty, 
Science \textbf{298}, 1372 (2002).

\bibitem{MBBBK}
J.~McKeever, A.~Boca, A.~D.~Boozer, 
J.~R.~Buck, and H.~J.~Kimble, 
Nature \textbf{425}, 268 (2003).

\bibitem{dutra}
S.~M.~Dutra,  
\textit{Cavity Quantum Electrodynamics} 
(Wiley-Interscience Publication, 
Ney York 2005).

\bibitem{HR06}
S.~Haroche and J.-M.~Raimond,  
\textit{Exploring the Quantum: 
Atoms, Cavities, and Photons} 
(Oxford University Press, 
Oxford, 2006).

\bibitem{H-R}
P.~Horak, G.~Hechenblaikner, K.~M.~Gheri, 
H.~Stecher, and H.~Ritsch,  
Phys. Rev. Lett. \textbf{79}, 4974 (1997).


\bibitem{H-K}
C.~J.~Hood, T.~W.~Lynn, A.~C.~Doherty, 
A.~S.~Parkins, and H.~J.~Kimble,  
Science \textbf{287}, 1447 (2000).

\bibitem{F-R}
T.~Fischer, P.~Maunz, P.~W.~H.~Pinkse, 
T.~Puppe, and G.~Rempe,  
Phys. Rev. Lett. \textbf{88}, 163002 (2002).

\bibitem{DR}
P.~Domokos and H.~Ritsch,  
J. Opt. Soc. Am. B \textbf{20}, 1098 (2003).

\bibitem{DR02}
P.~Domokos and H.~Ritsch,  
Phys. Rev. Lett. \textbf{25}, 253003 (2002).

\bibitem{GK}
C.~C.~Gerry and P.~L.~Knight, 
\textit{Introductory Quantum Optics} 
(Cambridge University Press, 
Cambridge, 2005). 

\bibitem{AEI}
A.~V.~Andreev, V.~I.~Emel'yanov, 
and Yu.~A.~Il'inski\v{\i}, 
\textit{Cooperative Effects in Optics} 
(Institute of Physics Publishing, 
Bristol, 1993). 


\bibitem{PL}
J.-S.~Peng and G.~X.~Li, 
\textit{Introduction to Modern Quantum Optics} 
(World Scientific, Singapore, 1998).


\bibitem{preparata}
G.~Preparata,  
\textit{QED Coherence in Matter} 
(World Scientific, Singapore, 1995).

\bibitem{enz}
C.~P.~Enz,  
Helv. Phys. Acta \textbf{70}, 141 (1997).

\bibitem{hir01}
M.~Hirokawa,  
Rev. Math. Phys. \textbf{13}, 221 (2001).

\bibitem{hir02}
M.~Hirokawa,  
Phys. Lett. A \textbf{294}, 13 (2002).


\bibitem{S-K99}
J.~Stenger, S.~Inouye, D.~M.~Stamper-Kurn, 
A.~P.Chikkatur, D.~E.Pritchard, and W.~Ketterle,  
Appl. Phys. B \textbf{69}, 347 (1999).

\bibitem{I-K}
S.~Inouye, A.~P.~Chikkatur, D.~M.~Stamper-Kurn, 
J.~Steger, D.~E.~Pritchard, and W.~Ketterle,  
Science \textbf{285}, 571 (1999).

\bibitem{S-K}
D.~Schneble, Y.~Torii, M.~Boyd, 
E.~W.~Streed, D.~E.~Pritchard, and W.~Ketterle,  
Science \textbf{300}, 475 (2003).

\bibitem{F-B}
L.~Fallani, C.~Fort, N.~Piovella, 
M.~Cola, F.~S.~Cataliotti, M.~Inguscio, 
and R.~Bonifacia,  
Phys. Rev. A \textbf{71}, 033612 (2005).

\bibitem{PVZ} 
J.~V.~Pul\'{e}, A.~F.~Verbeure, 
and V.~A.~Zagrebnov, 
J. Phys. A: Math. Gen. \textbf{38}, 5173 (2005). 

\bibitem{PVZ04} 
J.~V.~Pul\'{e}, A.~F.~Verbeure, 
and V.~A.~Zagrebnov, 
J. Phys. A: Math. Gen. \textbf{37}, L321 (2004). 

\bibitem{PVZ05} 
J.~V.~Pul\'{e}, A.~F.~Verbeure, 
and V.~A.~Zagrebnov, 
J. Stat. Phys. \textbf{119}, 309 (2005). 


\bibitem{hir-iumj}
M.~Hirokawa, 
to appear in Indiana Univ. Math. J. 


\bibitem{dicke}
R.~H.~Dicke, 
Phys. Rev. \textbf{93}, 99 (1954). 


\bibitem{PW} 
A.~Parmeggiani and M.~Wakayama, 
Forum Math. \textbf{14}, 539 (2002); 
\textit{ibid}. \textbf{14}, 669 (2002); 
\textit{ibid}. \textbf{15}, 955 (2003).   

\bibitem{NNW} 
K.~Nagatou, M.~T.~Nakao, 
and M.~Wakayama,  
Numer. Funct. Analy. Optim. \textbf{23}, 633 (2003). 

\bibitem{parmeggiani} 
A.~Parmeggiani, 
Kyushu J. Math. \textbf{58}, 277 (2004); 
Comm. Math. Phys. \textbf{279}, 285 (2008); 
\textit{Introduction to the spectral theory of non-commutative 
harmonic oscillators} 
(COE Lecture Note vol.8, Kyushu University, 
The 21st Century COE Program ``DMHF'', 
Fukuoka, 2008).  

\bibitem{IW} 
T.~Ichinose and M.~Wakayama,  
Commun. Math. Phys. \textbf{256}, 697 (2005); 
Rep. Math. Phys. \textbf{59}, 421 (2007).  


\bibitem{HL}
K.~Hepp and E.~H.~Lieb, 
Ann. Phys. (N.Y.) \textbf{76}, 360 (1973); 
Phys. Rev. A \textbf{8}, 2517 (1973); 
Helv. Phys, Acta \textbf{46} 573 (1973). 

\bibitem{SZ}
M.~O.~Scully and M.~S.~Zubairy, 
\textit{Quantum Optics} 
(Cambridge Univ. Press, 2006).


\bibitem{JC}
E.~T.~Jaynes and F.~W.~Cummings,  
Proc. IEEE \textbf{51}, 89 (1963).



\bibitem{MvdS}
H.~J.~Metcalf and P.~van der Straten, 
\textit{Laser Cooling and Trapping} 
(Springer-Verlag, New York, 1999). 

\bibitem{B-C-T}
F.~Bardou,~J.-P.~Bouchaud, 
A.~Aspect, and C.~Cohen-Tannoudji, 
\textit{L\'{e}vy Statistics and Laser Cooling} 
(Cambridge University Press, 
Cambridge, 2002). 


\bibitem{kato} 
T.~Kato, 
\textit{Perturbation theory for linear operators} 
(Springer-Verlag, New York, 1995). 

\bibitem{rs4} 
M.~Reed and B.~Simon,  
\textit{Methods of modern Mathematical Physics IV. 
Analysis of Operators} 
(Academic Press, San Diego, 1978). 


\bibitem{ME} 
P.~W.~Milonni and J.~H.~Eberly,  
\textit{Lasers} 
(Wiley Interscience Publication, New York, 1988). 


\bibitem{BR} 
S.~M.~Barnett and P.~M.~Radmore,  
\textit{Methods in Theoretical Quantum Optics} 
(Oxford University Press, 
Oxford, 2002). 




\bibitem{TRK}
R.~J.~Thompson, G.~Rempe, and H.~J.~Kimble,  
Phys. Rev. Lett. \textbf{68}, 1132 (1992).


\bibitem{B-H1}
M.~Brune, P.~Nussenzveig, F.~Schmidt-Kaler, 
F.~Bernardot, A.~Maali, J.~M.Raimond, 
and S.~Haroche,  
Phys. Rev. Lett. \textbf{72}, 3339 (1994).

\bibitem{B-H2}
M.~Brune, F.~Schmidt-Kaler, A.~Maali, 
J.~Dreyer, E.~Hagley, J.~M.Raimond, 
and S.~Haroche,  
Phys. Rev. Lett. \textbf{76}, 1800 (1996).

\bibitem{C-F}
J.~J.~Childs, K.~An, M.~S.~Otteson, 
R.~R.~Dasari, and M.~S.~Feld,  
Phys. Rev. Lett. \textbf{77}, 2901 (1996).

\bibitem{grangier}
P.~Grangier,  
Science \textbf{281}, 56 (1998).

\bibitem{NCA}
H.~Nha, Y.-T.~Chough, and K.~An,  
J. Korean Phys. Soc. \textbf{37}, 693 (2000).

\bibitem{IHB}
W.~T.~M.~Irvine, K.~Hennessy, 
and D.~Bouwmeester,  
Phys. Rev. Lett. \textbf{96}, 057405 (2006).

\end{thebibliography}
\end{document}